# Non-thermal Radio Emission from Massive Protostars in the SARAO MeerKAT Galactic Plane Survey

W.O Obonyo,[1,2]★ M.G Hoare[3], S.L Lumsden[3], M.A Thompson[3], J. O. Chibueze[1], W. D. Cotton[4,5], A. Rigby[3], P. Leto[6], C. Trigilio[6] G. M. Williams[7]

[1]*Department of Mathematical Sciences, University of South Africa, Cnr Christian de Wet Rd and Pioneer Avenue, Florida Park, 1709, Roodepoort, South Africa.*
[2]*Department of Astronomy and Space Science, The Technical University of Kenya, P.O. Box 52428 - 00200, Nairobi- Kenya.*
[3]*School of Physics and Astronomy, The University of Leeds, Woodhouse Lane, Leeds LS2 9JT, United Kingdom.*
[4]*National Radio Astronomy Observatory, 520 Edgemont Road, Charlottesville, VA 22903, USA*
[5]*South African Radio Astronomy Observatory (SARAO), 2 Fir Street, Black River Park, Observatory, Cape Town 7925, South Africa*
[6]*INAF – Osservatorio Astrofisico di Catania, Via S. Sofia 78, I-95123 Catania, Italy*
[7]*Department of Physics, Aberystwyth University, Ceredigion, Cymru, SY23 3BZ, UK*

## ABSTRACT

We present an investigation of the L-band emission from known massive young stellar objects (MYSOs) in the SARAO MeerKAT Galactic Plane Survey to search for non-thermal radio emitters in the sample. A total of 398 massive protostars, identified from the Red MSX Source (RMS) survey, are located within the survey region. Among these, 162 fields that host the protostars are isolated from nearby bright HII regions, allowing for the study of any ionized jets present. Seventy-one of these fields have jets with five-sigma detections or higher, corresponding to a detection rate of 44%. The MeerKAT fluxes of the detections, together with the upper limits of the non-detections and any other fluxes from previous observations, were used to estimate the spectral indices of the jets, and to search for the presence of non-thermal radiation. In cases where a source manifests as single in a given observation but is resolved into multiple components in observations of higher resolutions, the sum of the fluxes of the resolved components was used in estimating the indices. Any effects from missing flux in higher-resolution observations were incorporated into the index uncertainties. The spectral indices of the sample show that at least 50% of the jets emit non-thermal radiation. Additionally, the spectral energy distribution (SED) of some of the sources, as well as their radio luminosities exhibit evidence of non-thermal emission, especially in extended sources.

## 1 INTRODUCTION

The formation of massive stars ($M > 8\,M_{\odot}$) remains an active area of research, with a primary challenge of explaining how accreted materials overcome the strong radiation pressure from a forming massive protostar (Kahn 1974; Wolfire & Cassinelli 1987; Yorke & Sonnhalter 2002; Krumholz et al. 2005; Hosokawa & Omukai 2009). Various theories have been proposed to address this challenge, for example, varying dust-to-gas ratio (Wolfire & Cassinelli 1987), accretion at higher rates (McKee & Tan 2003; Hosokawa & Omukai 2009), disk-fed accretion and release of radiation pressure through outflow cavities (Krumholz et al. 2005; Kuiper et al. 2010).

Massive protostars undergo a phase in their evolution where they drive out jets (Guzmán et al. 2012; Moscadelli et al. 2016). While this phase is short-lived, it holds considerable significance as jets are considered the indirect indicators of disk-fed accretion (Kuiper et al. 2015). Kuiper et al. (2015), for example, shows that a fraction of the accreted mass is ejected through polar jets, directly linking the jet activity to the accretion process. The jet phase lies between the radio-quiet infrared dark clouds (IRDCs; Rathborne et al. 2006) and the radio-bright ultra-compact HII regions (UCHII; Churchwell 2002), and are thus classified as 'radio weak'. The jets are associated with cores that harbour the forming protostars, and in some cases, lobes that are mostly separated from the cores in high-resolution observations. The cores radiate free-free (thermal) radio continuum emission with characteristic positive spectral indices and have radio luminosities that lie in the range $0.1 \lesssim L_{\rm rad} \lesssim 100.0\,{\rm mJy\,kpc^2}$ (Anglada et al. 2018; Purser et al. 2016; Obonyo et al. 2019). In addition, some of the jets have radio lobes which may be thermal ($\alpha > -0.1$) or non-thermal with a typical spectral index $\alpha \simeq -0.6$ (Purser et al. 2016, Obonyo et al. 2019).

Previously, only the cores were detected since most observations were done at higher frequencies $\nu \geq 5\,$GHz, and lower sensitivities. However, the discovery of the polarised non-thermal jet in HH 80-81 (Carrasco-González et al. 2010), a massive young stellar object (MYSO), renewed interest in the magnetic properties of massive protostellar jets. Furthermore, previous studies primarily focused on targeted observations of nearby objects. The upgrade of both the Australia Telescope Compact Array (ATCA) and Jansky Very Large Array (JVLA), and now the construction of the MeerKAT, have enabled the study of these objects in large numbers and with high sensitivities despite their weak radio emission. Indeed, studies by Purser et al. (2016), Guzmán et al. (2012), Moscadelli et al. (2016) and Purser et al. (2021) all confirmed that ionized jets are common in massive protostars. Purser et al. (2016), for instance, classified 57% of their sample as jets. Even though these studies were conducted using the Australia Telescope Compact Array (ATCA) and Jansky Very Large Array (JVLA) at higher frequency bands ($\nu \geq 5$ GHz) where synchrotron emission is weak, Purser et al. (2016) demonstrated that approximately 40% of jet-driving MYSOs have non-thermal lobes.

★ E-mail: willice.obonyo@tukenya.ac.ke

The model of a massive protostar by Hosokawa & Omukai (2009) predicts a convective phase that can result in radio-emitting, collimated magneto-hydrodynamics (MHD) jets comparable to those driven by low mass protostars (Shang et al. 2004) through an interaction between stellar and disk magnetic fields (Shu et al. 1994). Additionally, the model proposed by Blandford & Payne (1982) suggests the possibility of driving astrophysical jets solely with magnetic fields of the accretion disc.

The SARAO[1] MeerKAT Galactic Plane Survey (Goedhart et al. 2024), conducted at 1.3 GHz, thus presents an opportunity for the search and study of the nature of the emission radiated by jet-driving MYSOs, especially their non-thermal radio emission. Its high sensitivity and capacity to detect extended emission make it suitable for identifying the extent of the weak, diffuse emission typical of lobes and bow shocks associated with jets. Combining this survey with previous radio and infrared (IR) observations, e.g, Lumsden et al. (2013) can enhance our understanding of both the small and large-scale features of these jets. The data will also be used to estimate the spectral indices of the MYSOs and identify unresolved, potential jet drivers for future follow-up observations at higher resolution.

## 2 OBSERVATION AND DATA

### 2.1 MeerKAT Data

The primary data used in this study were taken from the SARAO MeerKAT Galactic Plane Survey (Goedhart et al. 2024), conducted between 21$^{st}$ July 2018 and 14$^{th}$ March 2020. The survey imaged ~480 deg$^2$ of the Galactic Plane that lies between the longitudes; $2° < l < 60°$ and $255° < l < 358°$, and latitudes; $-1.5° < b < 1.5°$ with a spatial resolution $\theta \simeq 8''$ at 1.3 GHz. MeerKAT's shortest baseline is 29 m (Brederode & van den Heever 2017), translating to the largest angular size of 24′ at 20 cm. Its primary beam and highest sensitivity at the wavelength of the survey are $\theta_{PB} \simeq 50'$ and $\sim 10\,\mu$Jy beam$^{-1}$ respectively. Regions associated with diffuse emission, on the other hand, manifested lower sensitivities. In addition, to the regions of emission which can obscure faint compact sources, the imaging limitations of interferometers can produce artifacts near bright extended emission, both positive and negative, which can also hide fainter sources. A detailed description of how the data were processed is given in Goedhart et al. (2024).

## 3 IDENTIFICATION OF SAMPLE

The massive protostars in this study were selected from the Red MSX Source (RMS[2]) Survey catalogue (Lumsden et al. 2013). The catalogue lists 398 massive protostars located within the region observed by the MeerKAT Galactic Plane Survey (GPS). Most of the protostars have corresponding radio observations at different frequencies e.g. Urquhart et al. (2007) and Purcell et al. (2013). The fields of all the protostars were visually inspected to eliminate non-detections, e.g. the field of G023.8176+00.3841 shown on the top panel of Figure 1. Similarly, objects whose emission are heavily affected by nearby HII regions, e.g. G011.9920−00.2731 shown on the lower panel of Figure 1, were also excluded. While the contour map of G011.9920−00.2731's field suggests the presence of two weakly detected radio sources, their emissions are blended with the diffuse emission in the field and could not be separated.

A total of 162 fields, comprising about 41% of the total, exhibited minimal effects from nearby bright sources. Among these, 71 fields showed significant detections of massive protostars with a signal-noise ratio (SNR) of 5 or greater - see Table A1. The infrared fields of the 71 sources were also examined to search for evidence of outflows by visually inspecting the distribution of dust emission at infrared wavelengths (Ohlendorf et al. 2012). Infrared Array Camera (IRAC) images from the Galactic Legacy Infrared Mid-Plane Survey Extraordinaire (GLIMPSE; Churchwell & GLIMPSE Team 2001) survey with a wavelength range of $3.6 - 8\,\mu$m emission were inspected. All 71 fields displayed evidence of jet-like features.

Further, a check on the match between the radio sources with their RMS counterparts was done by calculating the offsets between their positions in the MeerKAT observation and Lumsden et al. (2013) (see figure 2), even though the positions in Lumsden et al. (2013) were derived from both infrared (Skrutskie et al. 2006, Churchwell et al. 2009, Benjamin et al. 2003) and radio (Purcell et al. 2013) observations. The minimum angular separation, $\theta_{min\,sep}$, beyond which unresolved objects were considered separate, was estimated by combining the uncertainties in the positions of the observations (Kwok et al. 2008; Goedhart et al. 2024), in quadrature, according to Equation 1 where PA $\simeq \frac{\mathrm{FWHM}}{2} \times$ SNR (Condon et al. 1998) is the position accuracy of observation.

$$\theta_{\mathrm{min\,sep}} \simeq \sqrt{\mathrm{PA_1}^2 + \mathrm{PA_2}^2} \quad (1)$$

The average uncertainty in the separation distance of unresolved radio sources with an SNR of 5 was approximately 4″. Consequently, sources with angular displacements exceeding 4″ from the position of their infrared counterparts were considered separate. Since radio and infrared emissions might be tracing different regions of the MYSOs, sources shown on the upper panel of Figure 2 with separation distances greater than 4″ were examined at other frequencies to establish their nature and positions more accurately. For example, G032.0518-00.0902 is unresolved at 1.3 GHz but is known to radiate $2.12\,\mu$m $H_2$ emission (Cooper et al. 2013), which is a shock tracer and an indirect indicator of the presence of a jet. Such objects were retained regardless of the magnitude of their offsets. An example of a radio source eliminated due to a large offset with the IR image in the field is G286.2086+00.1694, with an offset of 7.53″ (see Figure 3). Indeed, Pitts et al. (2018) and Barnes et al. (2023) identified both sources as separate mid-infrared emitters. The range and average offsets of the 71 MYSOs selected were found to be 0.1−4.7″ and 1.6±0.9″ respectively, with the majority of the sources having an angular offset $\theta \sim 1.25''$ (see the upper panel of Figure 2). The largest offsets were seen in MYSOs with extended morphology in the MeerKAT observation, such as G035.1979-00.7427 and G059.7831+00.0648, where offsets $\theta > 2''$.

Typically, protostellar jets have cores that host the forming protostars and are largely co-located with infrared sources, and lobes that are normally a few arcseconds away from the cores. Thus, if the offset of an unresolved radio source was larger than four arcseconds but it lies along the axis of a known jet then it was retained as a potential lobe. For example, Navarete et al. (2015) detected six 2.12$\mu$m molecular hydrogen sources within the field of G019.8817-00.5347, three on either side of the MYSO, implying the presence of a bipolar outflow. MeerKAT also detected two radio emitters that are coincident with the molecular hydrogen sources and lie in the same orientation, suggesting that they are potential lobes (see Figure 4). The two radio lobes were thus retained even though their offsets are greater than 4″. It is unclear why the third lobe was not detected at 1.3 GHz, however, some lobes are weak thermal emitters, which could result in weak or undetectable radio emission.

[1] South African Radio Astronomy Observatory
[2] RMS website:http://rms.leeds.ac.uk

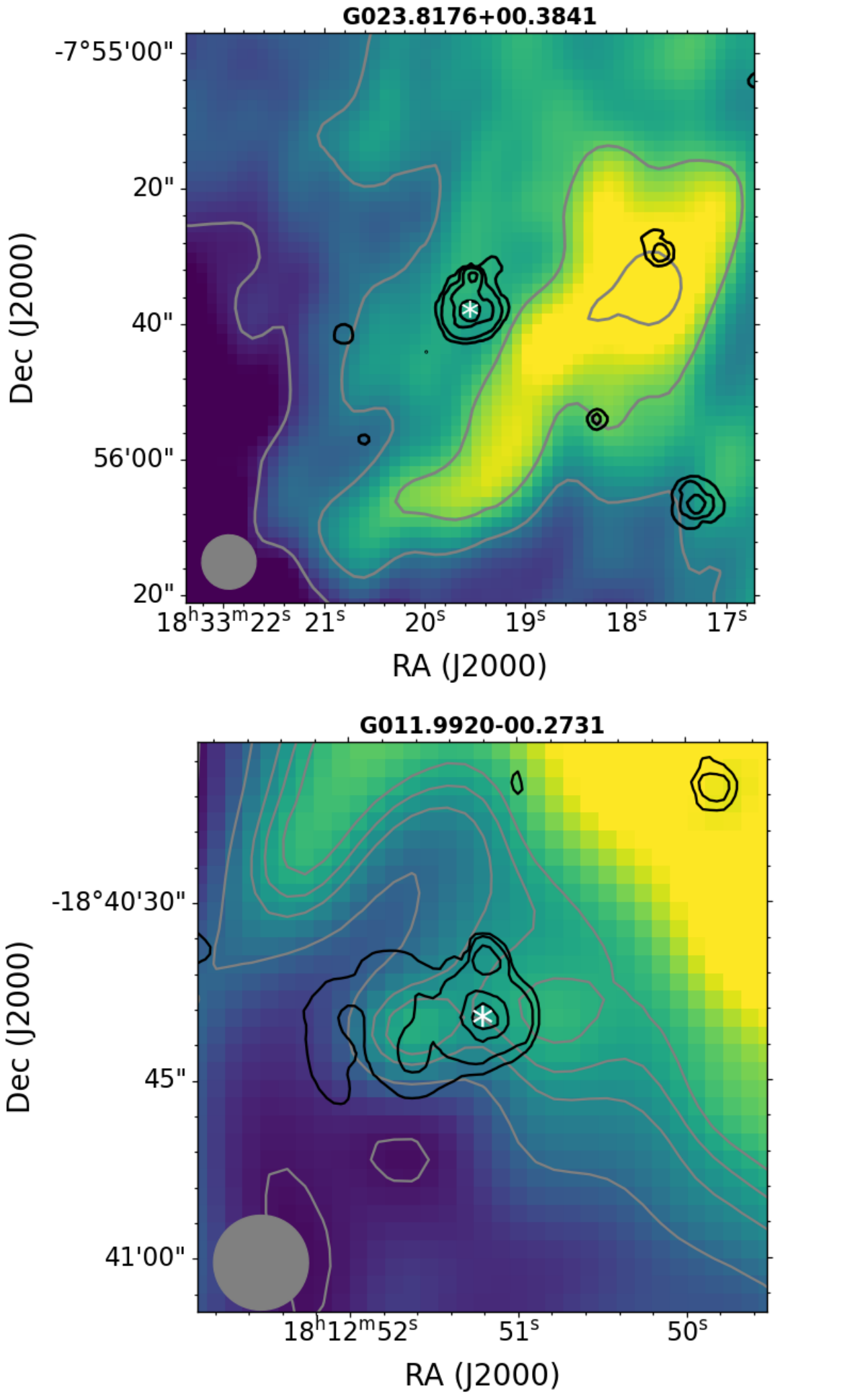

**Figure 1.** Top: A plot illustrating the non-detection in the field of G023.8176+00.3841, attributed to the strong emission from an extended nearby object, depicted by the colour map and grey contours (3, 11, 20, and 30$\sigma$) of 1.3 GHz emission. The rms noise of the field, $\sigma$ = 76, $\mu$Jy/beam. Bottom: A plot showing how the radio emission from G011.9920−00.2731 is heavily contaminated by emission from a nearby HII region. The grey and black contours represent its radio and infrared emission at 1.3 GHz and 8 $\mu$m, respectively. The infrared positions of the MYSOs are indicated by white asterisks (*) marks. The beams are shown in the lower left corner.

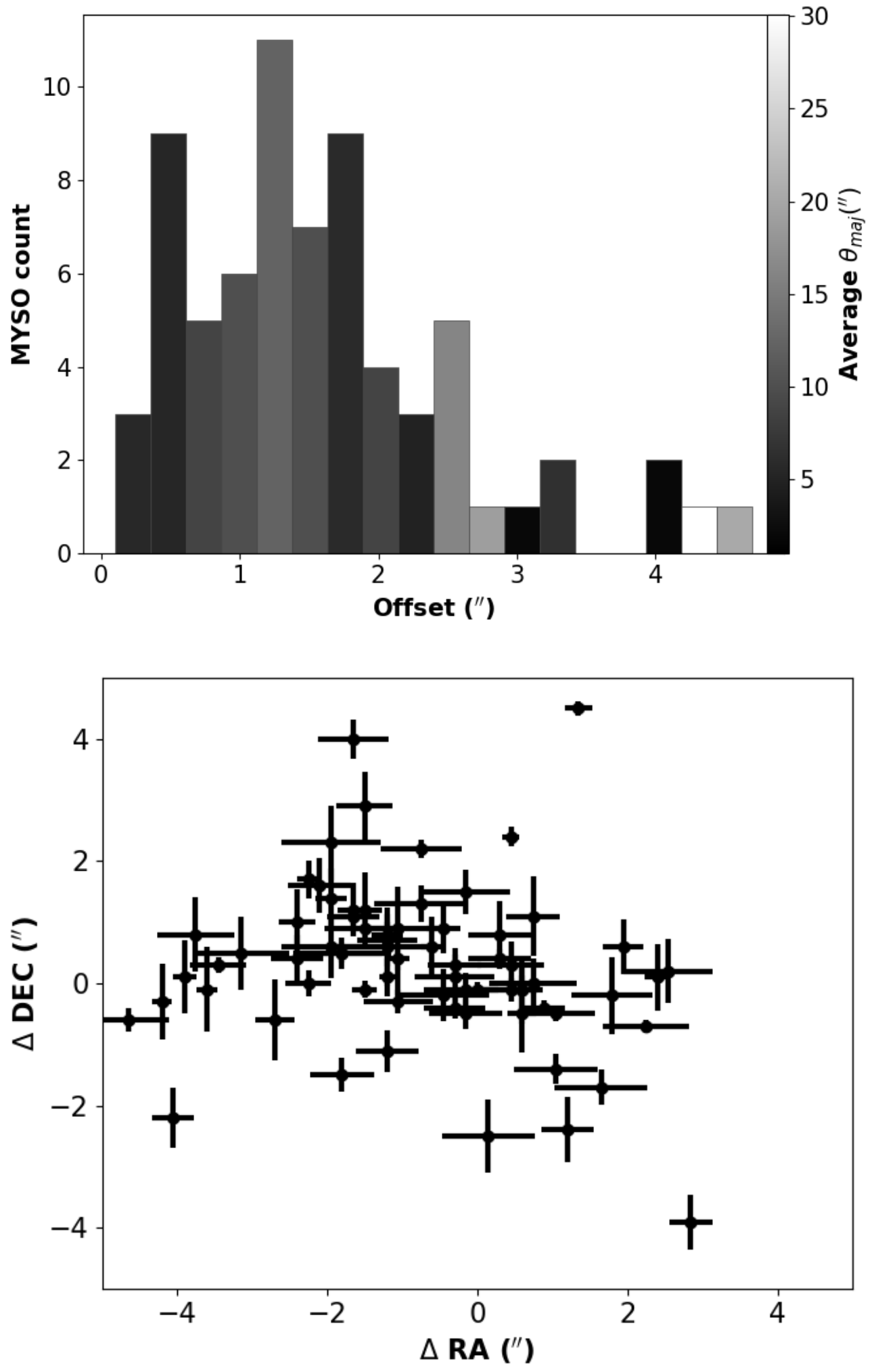

**Figure 2.** Top: A histogram of the angular separation between the positions of the sources in Lumsden et al. (2013) and MeerKAT observations. Bottom: The offsets in declination and right ascension.

Finally, the higher spatial resolution of the GLIMPSE observation, < 2″ (Churchwell et al. 2009), meant that two IR sources with an angular separation $\theta$ < 8″ could not be resolved in the MeerKAT observation. Such sources, e.g, G042.0977+00.3521A, G042.0977+00.3521B and G260.9252+00.1149A&B (see Figure 5), were also eliminated as their fluxes and positions could not be measured in the MeerKAT observation.

## 4 RADIO PROPERTIES OF MASSIVE PROTOSTARS IN THE SARAO MEERKAT GPS

The integrated fluxes and angular sizes of the MYSOs detected in the survey were determined by fitting two-dimensional Gaussian functions to their emission. The emission fit well with the Gaussian model in all the cases. Approximately 70% of the sources have integrated fluxes $S_\nu \leq 1.5$ mJy, with the faintest and brightest sources in the sample measuring 0.04±0.01 mJy and 51.8±0.8 mJy respectively (see Figure 6). For non-detections, an upper limit on the fluxes was set at 3$\sigma$, where $\sigma$ is the root-mean-square (rms) noise of the field. The radio luminosities of the sources at 1.3 GHz were calculated using the distances provided in Lumsden et al. (2013) and found to be consistent with those of jets driven by massive protostars ($L_{bol} \simeq 2000\,L_{\odot}$) at 1.5 GHz, i.e, $L_{rad} \leq$80 mJykpc$^2$ (Obonyo et al. 2019).

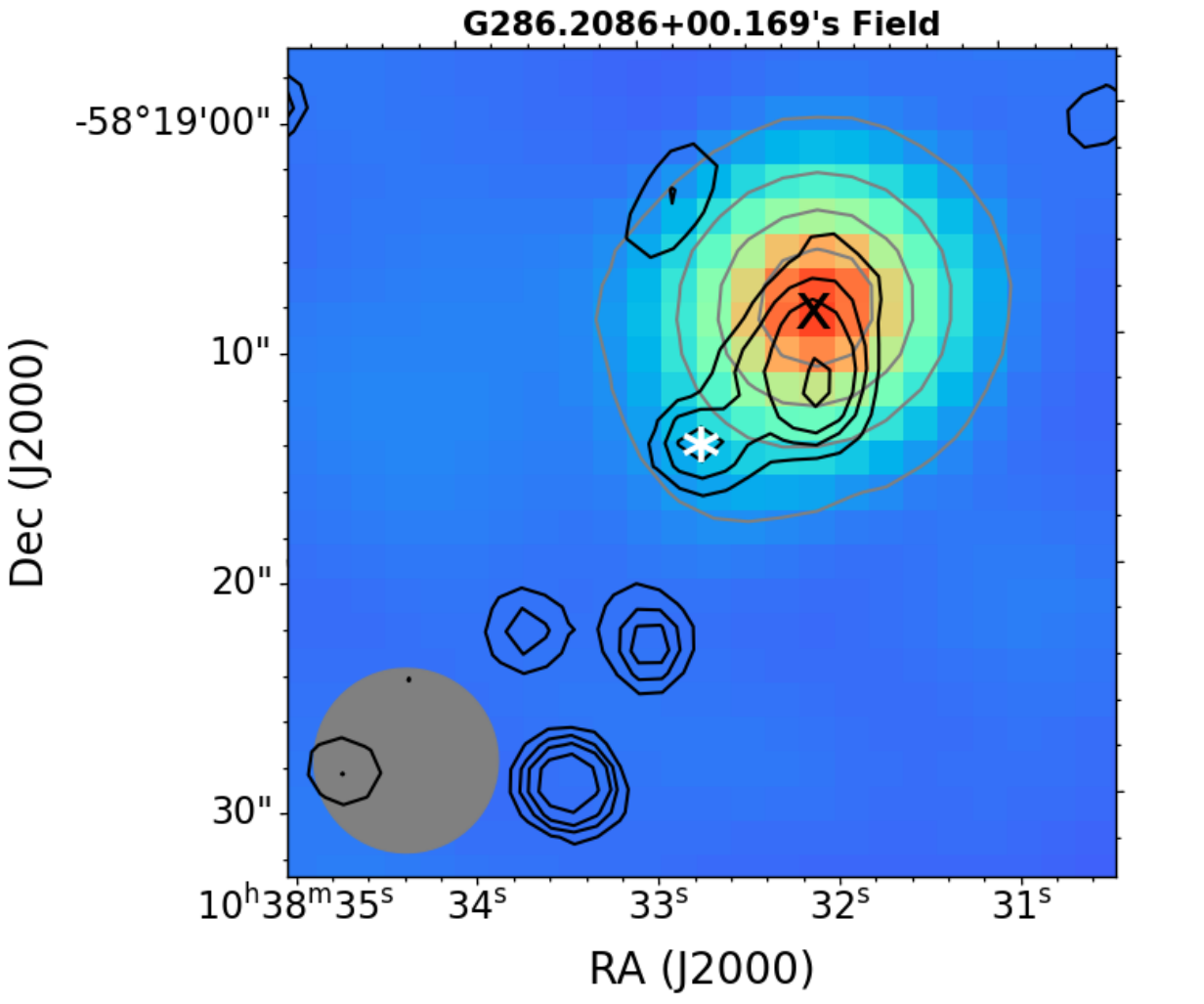

**Figure 3.** A plot showing one of the largest offsets between the position of the radio (marked with x) and infrared (marked with *) sources in G286.2086+00.1694. The separation between them is 7.53″. The emission from 2MASS Ks-band and MeerKAT 1.3 GHz are represented by the black and grey contours, respectively, with the MeerKAT contour levels set at 3, 11, 20, 30 times 36 $\mu$Jy/beam. The grey ellipse shows the synthesised beam of the MeerKAT observation.

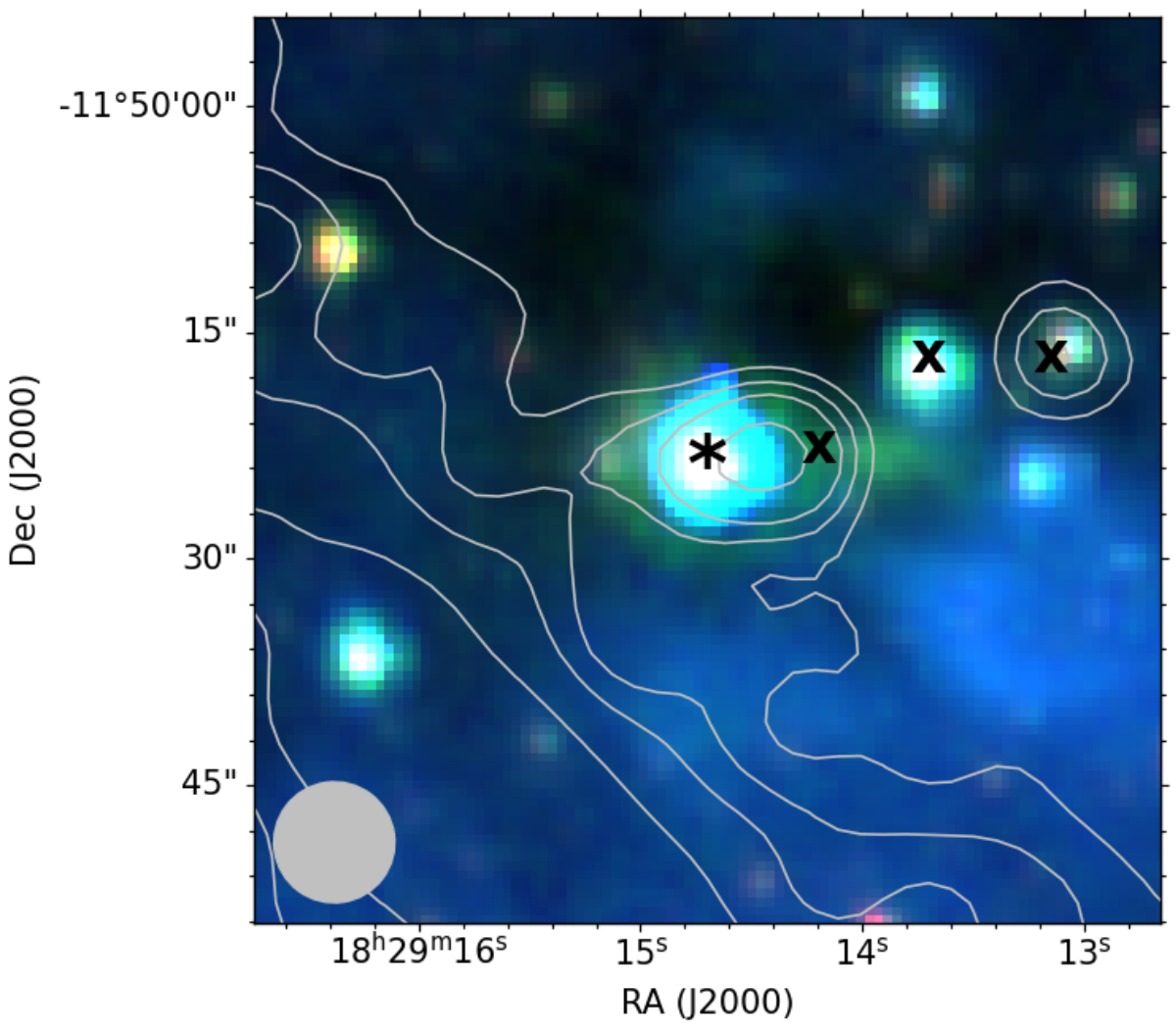

**Figure 4.** Three colour map of GLIMPSE 3.6 $\mu$m, 5.8 $\mu$m and 8.0 $\mu$m emissions within the field of G019.8817-00.5347. The grey contours represent MeerKAT's 1.3 GHz emission at levels 3,5,7 and 9$\sigma$. The asterisk sign represent the location of the MYSO (Lumsden et al. 2013) while the cross (x) signs are the locations of 2.12$\mu$m molecular hydrogen emission (Navarete et al. 2015). The grey ellipse shows the synthesised beam of the MeerKAT observation.

The deconvolved angular sizes of both major and minor axes of the protostars exhibit a wide range, from 2.5±0.5″ to 30.1±0.9″ for major axes, and from 0.9±0.3″ to 20.0±0.7″ for minor axes. The ratios of the major-to-minor axes of the sources were calculated to identify sources with extended and elliptical radio emission. The emission morphologies of massive protostellar cores are typically elliptical, with the major axes aligned with the poles of the jets (Anglada et al. 2018). A source was considered extended if it was larger than the beam and elliptical if its major axis was at least 1.2 times its minor axis (see Figure 7). The lower limit of the ratio of the major-to-minor axes of the sources was deduced from the largest observed opening angle, $\theta_o$, in a sample of massive protostellar jets, which was 80° in Purser et al. (2016). The positions of the sources in Figure 7 suggest that most of them are potential jet drivers. Indeed, even the size of a massive protostellar jet that is known to drive a jet and has lobes, G263.7759-00.4281 (Purser et al. 2016), was unresolved in the MeerKAT observation, and thus classified as a point source, implying that most, if not all, of the sources shown on the plot may be jet drivers.

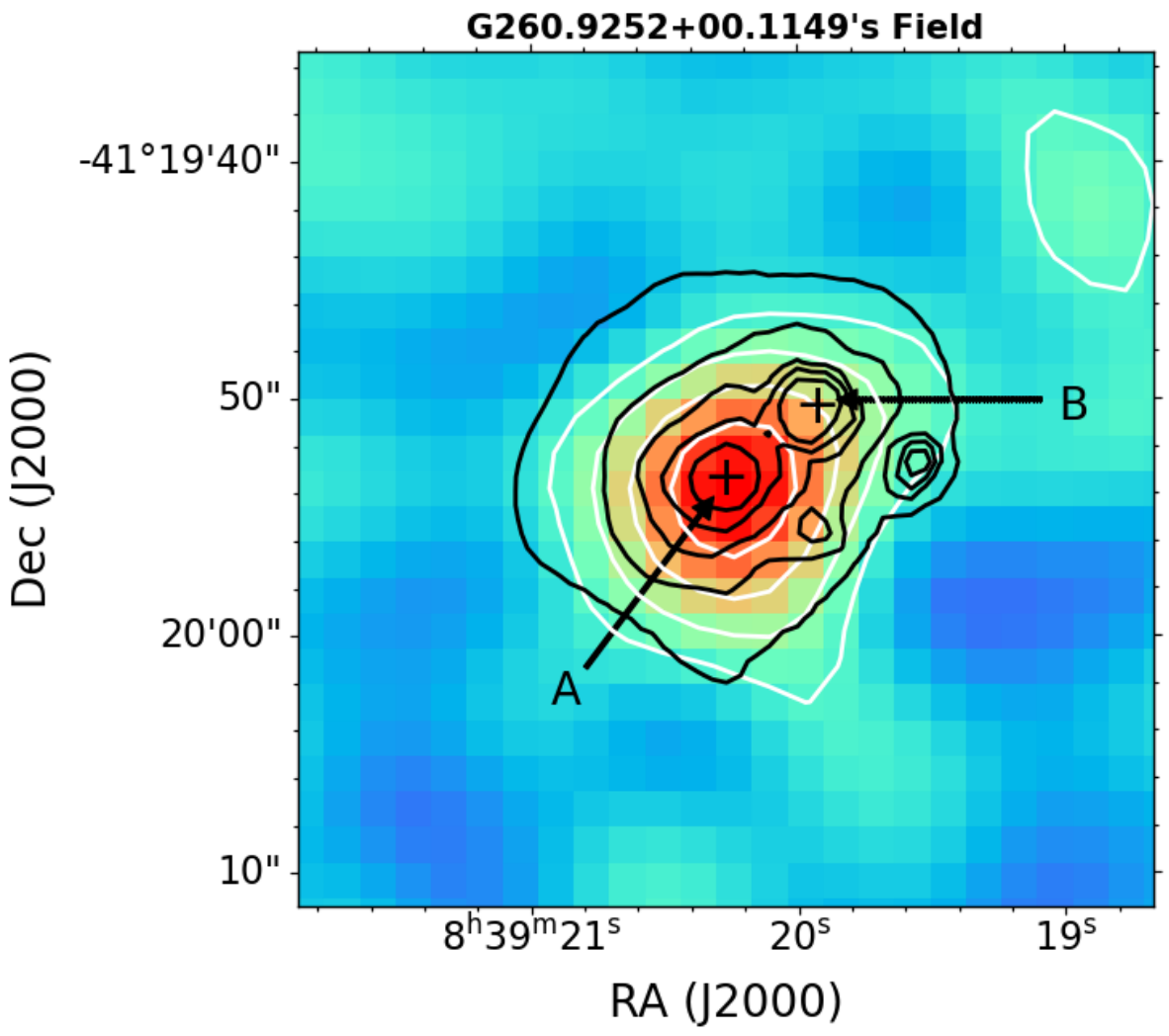

**Figure 5.** Colour map and white contours showing MeerKAT 1.3 GHz emission at levels 3,7,15,30 and 45$\sigma$, together with 8 $\mu$m contours shown in black. The plus signs represent the locations of the MYSOs (Lumsden et al. 2013).

### 4.1 Spectral Index Derived from Observations of Various UV-Coverages

This study utilizes observations of various uv-coverages to estimate the spectral indices of protostellar jets. Consequently, we conducted simulations of observations with resolutions and frequencies similar to those in the actual study. This aimed to estimate the proportion of flux that is recoverable or gets resolved out across the different frequency bands. To achieve this, we constructed a sky model representing a jet with flux density $S_\nu$ = 0.5 mJy at 5 GHz, opening angle $\theta_o$ ≃25° and angular dimensions of 2″ by 0.4″. The spectral index of the jet was 0.6, typical of ionized conical winds (Reynolds 1986). The sky model was used to predict the visibilities of the observations, with thermal noise added to the data to simulate real observing conditions. Synthetic observations of the jet were generated across five different frequencies: L-band to mimic MeerKAT observation and C-K bands to replicate JVLA high-resolution observations at its A-configuration. The resolutions of the synthetic observations at 1.5, 6.0, 10.0, 15.0 and 22.0 GHz were set at 8.0″, 0.33″, 0.2″, 0.13″, and 0.089″, respectively, resulting in synthetic UV-ranges of 0.4−34 k$\lambda$, 1.5−290 k$\lambda$, 5−480 k$\lambda$, 10−720 k$\lambda$, and 15−1050 k$\lambda$, respectively.

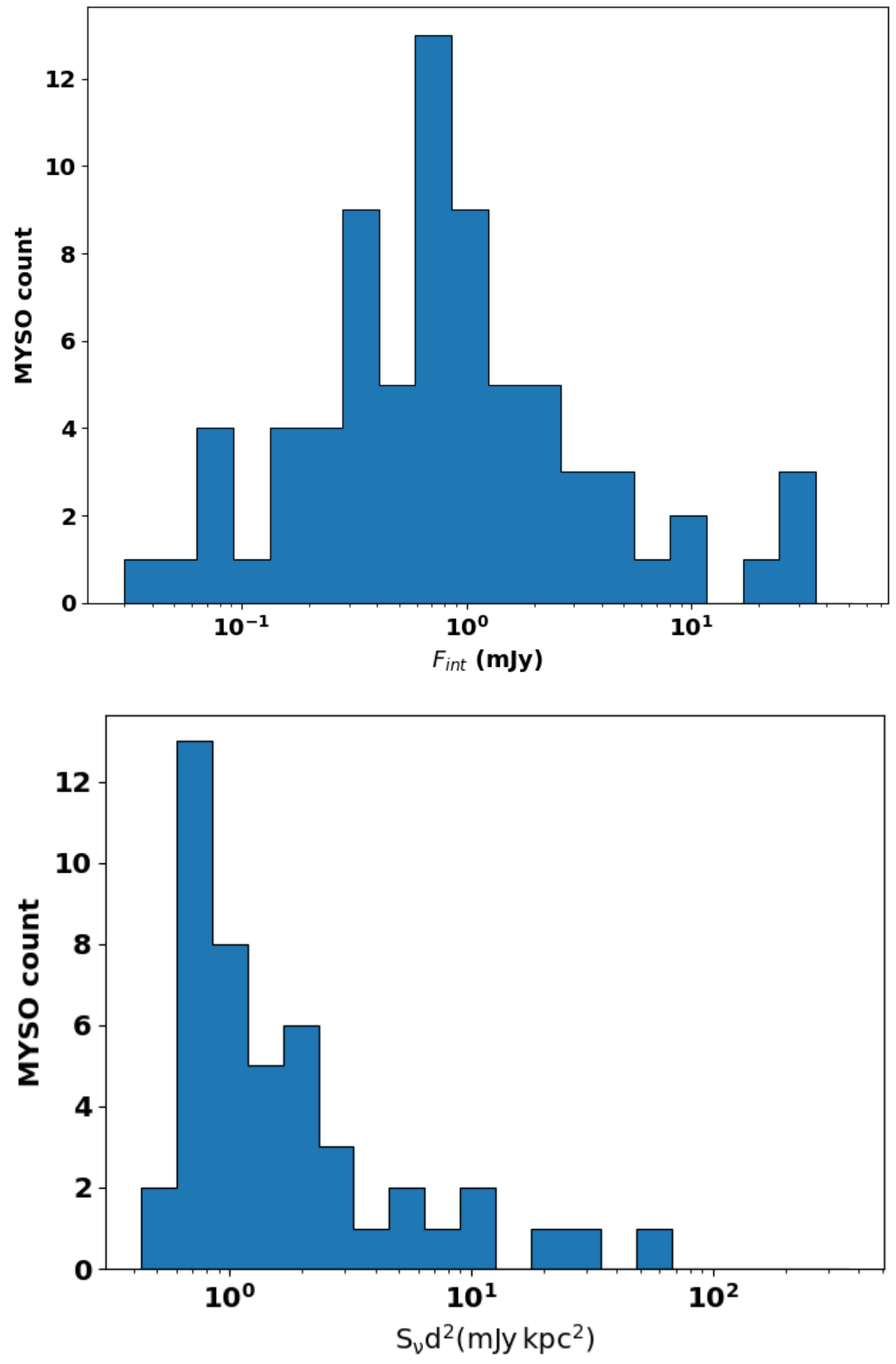

**Figure 6.** The distributions of the integrated fluxes (top) and radio luminosities (bottom) of the sample. The distances used in calculating the radio luminosities were taken from Lumsden et al. (2013).

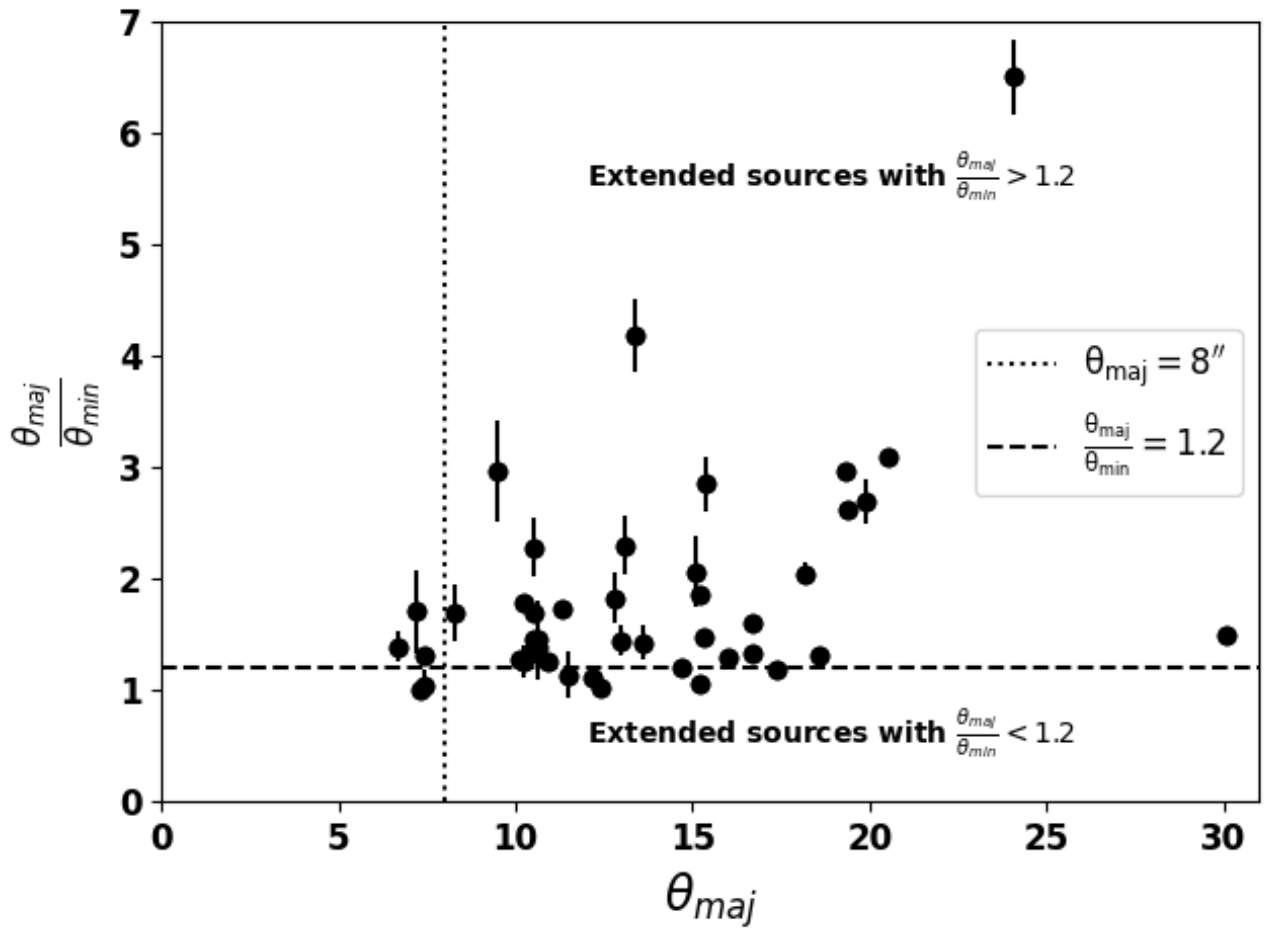

**Figure 7.** A plot of the ratio of major-to-minor axes vs major axes of the massive protostars. The ratio of the axes for most of the sources manifest jet properties.

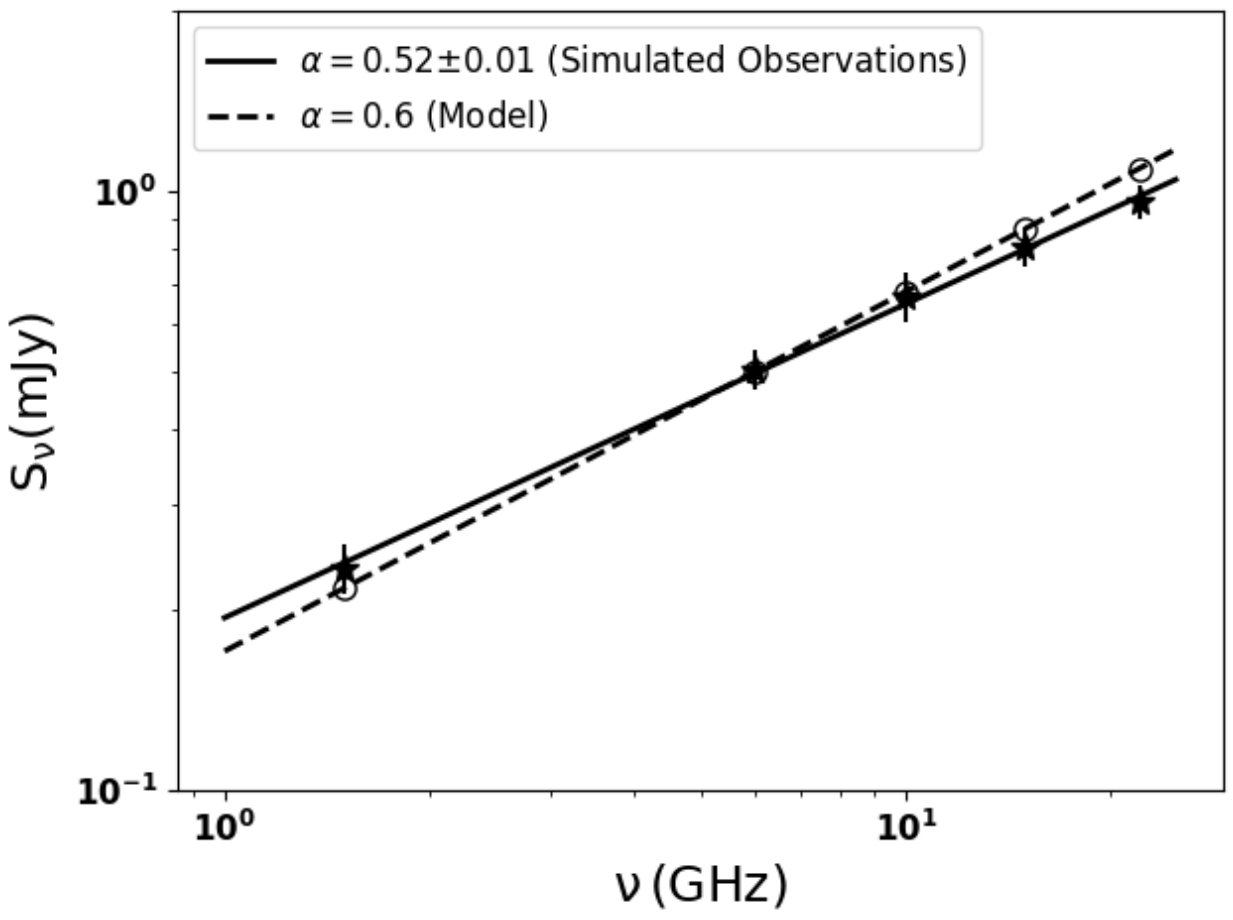

**Figure 8.** Fluxes for the model (depicted by empty circles) and synthetic observations (asterisks). The observed fluxes were derived by fitting a 2D Gaussian function on the images. The lines represent the Spectral Energy Distributions (SEDs) of the model and synthetic observation, obtained by the least squares method.

Figure 8 illustrates the fluxes and SEDs of both the model and synthetic observations of the jet. The fluxes recovered at L, C, X, Ku and K-bands were 108%, 101%, 98%, 93% and 88% of the model's flux, respectively, indicating an additional 8% of flux recovered at L-band while 12% of flux is resolved out at K-band. This leads to a decrease in the spectral index of the model by approximately 13%, an uncertainty that was added, in quadrature, while estimating the indices.

### 4.2 Sources with extended emission and/or radio lobes

Most of the studies of the extended morphologies of protostellar jets are conducted using data from observations of higher resolution where the lobes of the jets are well resolved and separated from the cores. In this observation, only three radio sources, G059.7831+00.0648, G343.1261-00.0623, and G345.4938+01.4677, have lobes that are separated from the cores. Nonetheless, their cores are known to harbour multiple radio components (Guzmán et al. 2010, Purser et al. 2016) along the jets. Thus, the sensitivity of the MeerKAT observation to extended emission allows us to explore the nature and extent of the extended emission that envelopes the components of such jets. Some of the sources with components at higher-frequency observations do not manifest any specific morphology at 1.3 GHz, however, a few display elongation along the poles of the jets in the MeerKAT observation. Moreover, spectral indexing was used to characterise their radiation, either as non-thermal (synchrotron) radiation or thermal (free-free) emission (Rodríguez-Kamenetzky et al. 2017, Obonyo et al. 2019). A detailed description of the elongated radio sources in the MeerKAT observation is given in the following sub-sections.

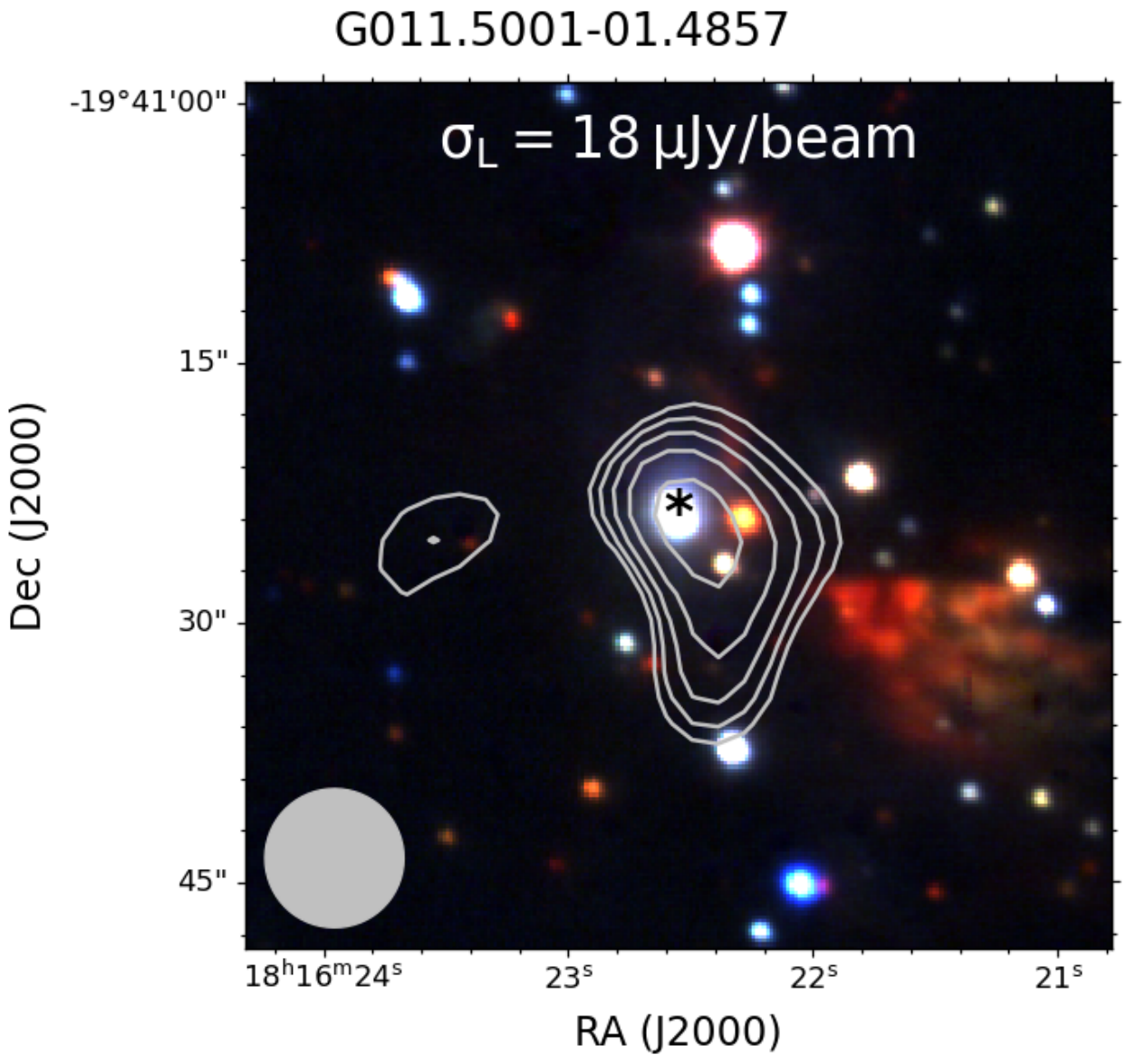

**Figure 9.** UKIDSS three colour image of G011.5001+01.4857's field. The silver contours of the radio emission at 1.3 GHz are plotted at 3,15,30,50 and 80$\sigma$. The asterisk sign represents the location of the main IR source. The grey ellipse shows the synthesised beam of the MeerKAT observation.

**4.2.0.1 G011.5001-01.4857:** G011.5001-01.4857 shows an extended emission of flux density 0.21±0.07 mJy at 1.3 GHz. It is largely aligned in a north-south direction (see Figure 9). Given that it encloses three infrared (IR) sources, the emission may be due to all or some of the sources with the lower end of the emission potentially due to outflow activity as there are no infrared sources towards the south of the emission. A compact radio source that is associated with the IR source was also detected at 8.4 GHz (van der Walt et al. 2003) with a flux density of 0.30±0.07 but not at 5 GHz (Urquhart et al. 2009) where the rms noise of the field is 0.15 mJy, giving it a spectral index of 0.19±0.27 between 1.3 and 8.4 GHz, hence thermal. Given that the radio source displays an extended morphology at 1.3 GHz and a compact structure at 8.4 GHz (van der Walt et al. 2003), its L-band emission may be from a core-lobe system while only the core is detected at 8.4 GHz.

**4.2.0.2 G017.6380+00.1566:** G017.6380+00.1566 displays an asymmetric extended emission in a SE-NW direction at 1.3 GHz, perhaps signifying radio extinction effects or Doppler boosting. Its 1.3 GHz radio emission of flux density 1.86±0.29 mJy encloses radiation from two radio sources, denoted as RS3 and RS4 in Menten & van der Tak (2004). The line joining RS3 and RS4 is aligned with the radio emission at 1.3 GHz i.e in a SE-NW direction as shown in Figure 10, implying that they could be a core-jet pair. The UKIDSS IR emission also has a similar orientation as that of the radio sources further signifying outflow in the same direction. A higher resolution 1.3 mm dust continuum map of the field (Maud et al. 2018), on the other hand, displays the presence of three cores that are closely located to the position of RS3, perhaps the driving source of the jet with RS4 as its lobe a jet lobe. The flux densities of G017.6380+00.1566

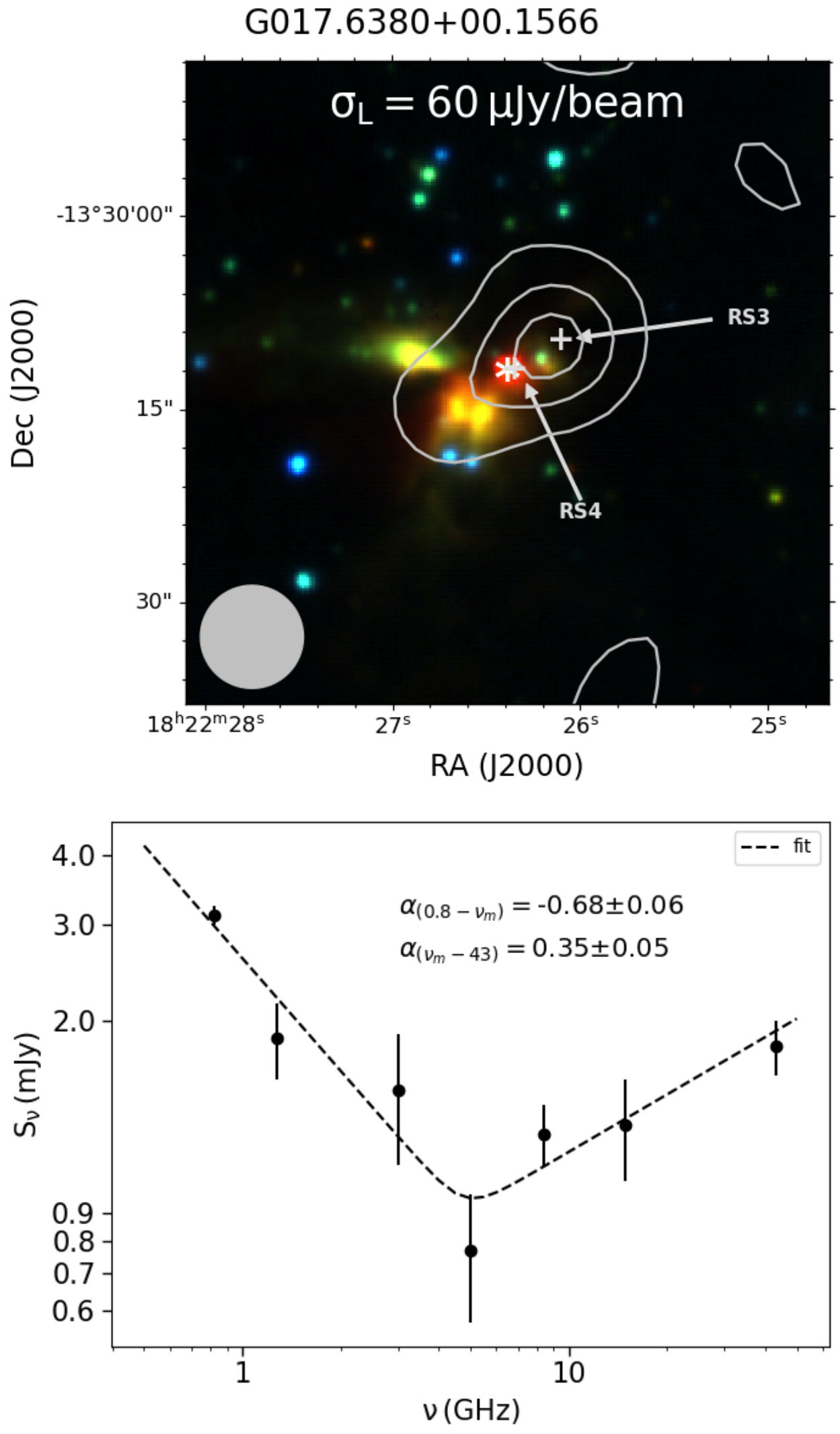

**Figure 10.** Top: UKIDSS three colour image of G017.6380+00.1566's field and grey contours of radio emission at 1.3 GHz (3,7 and 11$\sigma$). The two plus signs represent the locations of the radio sources detected at higher frequencies in Menten & van der Tak (2004) while the asterisk shows the position of the infrared source. The grey ellipse shows the synthesised beam of the MeerKAT observation. Bottom: SED of G017.6380+00.1566.

at 5.0, 8.4, 14.9 and 43.3 GHz are 0.77 ±0.2, 1.25±0.16, 1.30±0.27 and 1.8±0.2 mJy respectively (Menten & van der Tak 2004), giving it an overall positive spectral index of 0.03 ± 0.11 between 1.3 and 43.3 GHz, however, an inspection of the SED suggests that the radiation at 1.3 GHz is dominated by non-thermal emission (see Figure 10). Extrapolation of the SED of the thermal component of the emission to 1.3 GHz suggests that ~ 2.58 mJy (80%) of the emission at 1.3 GHz is non-thermal. Indeed the spectral index of the radio emission between L and C-bands $\alpha_{LC} = -0.68 \pm 0.06$ is typical of non-thermal emission.

**4.2.0.3 G035.1979-00.7427:** G035.1979-00.7427's flux density and deconvolved angular size at 1.3 GHz are 21.8±2.3 mJy and 19.8±2.2″ by 7.4±1.2″ respectively. Its emission at 1.3 GHz, shown in the upper panel of Figure 11, manifests an extended structure which resolves into multiple radio sources at higher resolution observations, e.g., nine at 5.8 GHz (Purser et al. 2021), eight at 5 GHz (Gibb et al. 2003), ten at 8.5 GHz (Gibb et al. 2003), sixteen at 15 GHz (Beltrán et al. 2016) and fourteen at 23 GHz (Beltrán et al. 2016). The sources are aligned in a N-S direction and spread across a region that is approximately 15″ by 5″. The MeerKAT flux together with the sum of the fluxes of the components of the jet in Purser et al. (2021), Gibb et al. (2003) and Beltrán et al. (2016) were used to plot the SED of the jet shown in the lower panel of Figure 11 and to estimate the spectral index of its emission. The SED shows a negative spectral index of -0.69±0.04 between 1.3 and 8.5 GHz and 0.67±0.10 between 8.5 and 23.1 GHz, implying that the jet emits a combination of thermal and non-thermal emission at all the radio bands with non-thermal emission dominating at 1.3 GHz while thermal emission is prominent at 23.1 GHz.

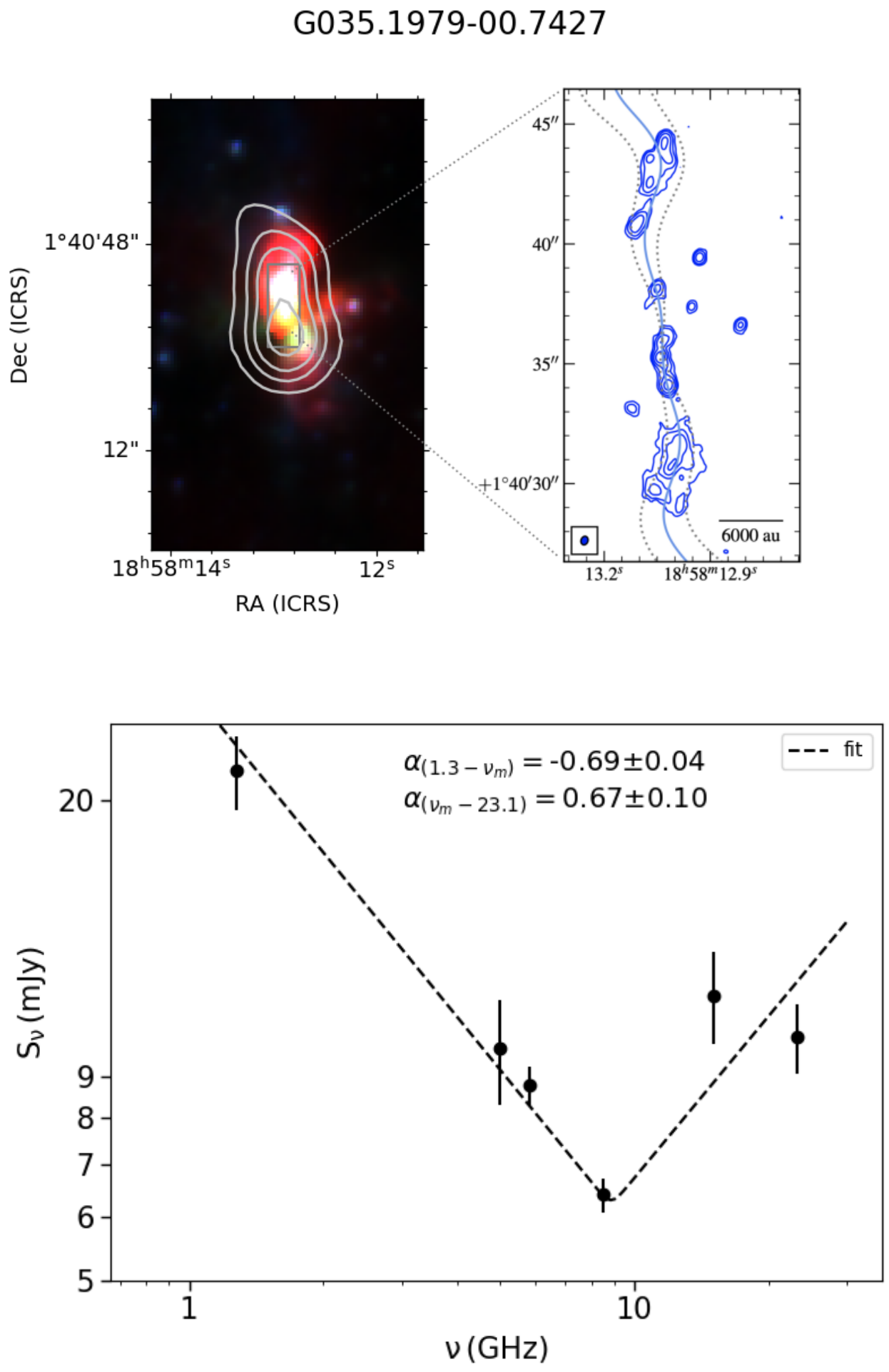

**Figure 11.** Top left: GLIMPSE three-colour image of G035.1979-00.7427 overlaid with contours of 1.3 GHz emission at levels 3,7,11 and 14$\sigma$. Top right: Blue contours of the source's emission at 5.0 GHz (Purser et al. 2021). Bottom: Spectral energy distribution of G035.1979-00.7427 between L- and K-bands.

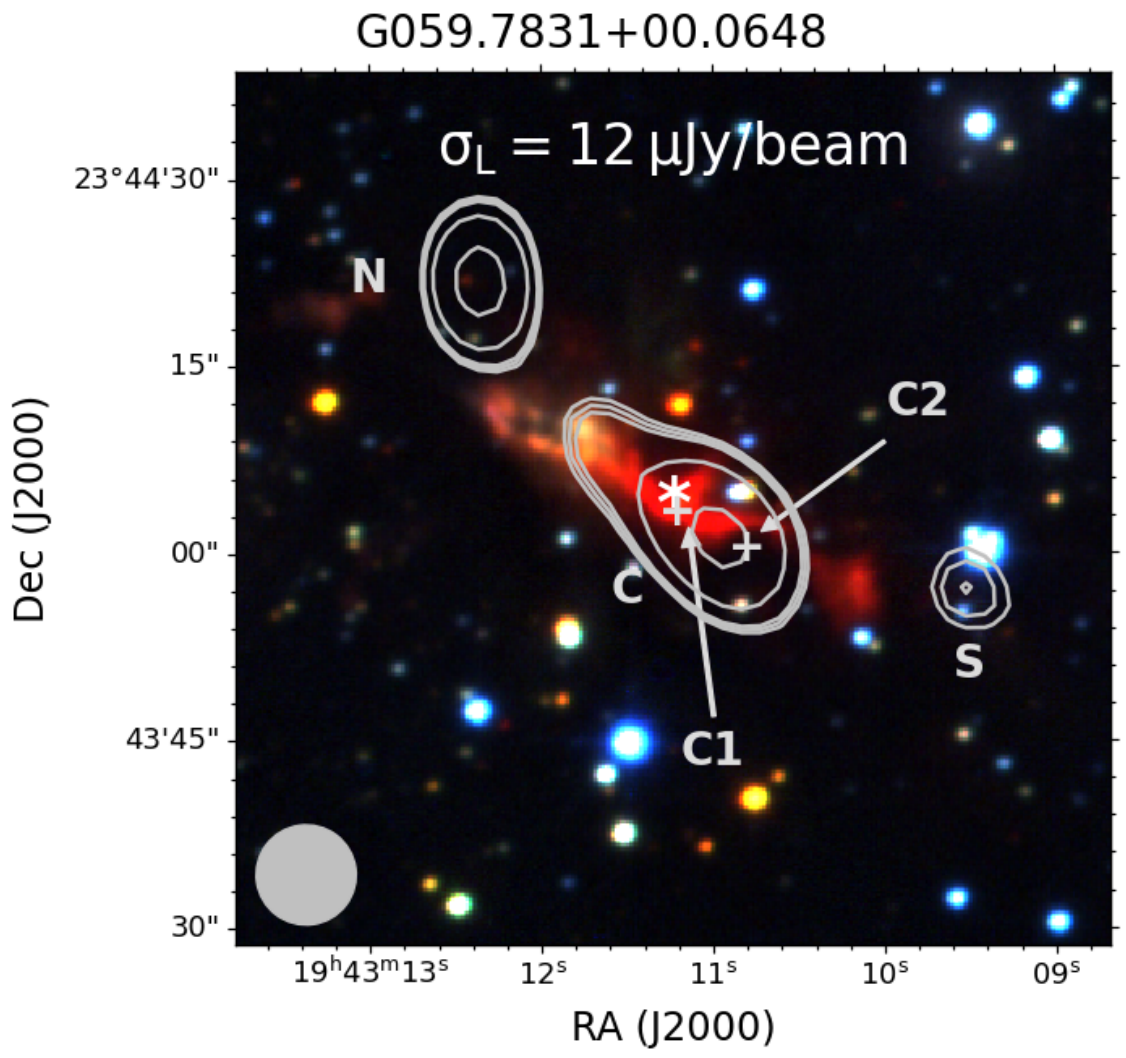

**Figure 12.** Three colour map of UKIDSS's J,H,K emissions in G059.7831+00.0648's field. The contours map for 1.3 GHz emission are of levels 3,5,7,20 and 50$\sigma$. The plus signs represent the location of radio sources detected by Carral et al. (1999) at 8.3 GHz. The asterisk next to C1 represents the location the infrared source. The grey ellipse shows the synthesised beam of the MeerKAT observation.

**4.2.0.4 G059.7831+00.0648:** Three radio sources that are aligned with the infrared emission of the field were detected in G059.7831+00.0648 at 1.3 GHz. The brightest source, C, which has a flux density of 0.43±0.09 mJy, is located between two other radio sources, N and S, whose flux densities are 0.26±0.08 mJy and 0.07±0.02 mJy respectively, perhaps its associated jet lobes. Carral et al. (1999) observed the field at 8.3 GHz and 43 GHz, detecting two radio sources at 8.3 GHz but not 43 GHz, where the rms was 1.5 mJy/beam. The two radio sources, marked C1 and C2 in Figure 12, are located within the region covered by C's emission. Their flux densities at 8.3 GHz are 0.7±0.14 and 0.30±0.06 mJy respectively. These fluxes, along with the flux of C at 1.3 GHz, were used to calculate the index of C, which is 0.29±0.16. An upper limit was also set on the indices of N and S using their 3$\sigma$ flux densities at 8.3 GHz and fluxes at 1.3 GHz, resulting in $\alpha < -0.7$ for N and $< 0.37$ for S, indicating that N is non-thermal while S is thermal. The orientation of UKIDSS, 2MASS, and GLIMPSE emission are comparable to that of the radio emission whose position angle, PA=48±3°, implying that the direction of the IR outflow cavity is similar to that of the jet. Similarly, $H_2$ emission (Beuther et al. 2003) in the field is aligned with radio emission, signifying the presence of shocks along the jet.

**4.2.0.5 G254.0548-00.0961:** G254.0548-00.0961, associated with IRAS 08159-3543, has an extended radio jet of position angle PA = 44±4° and an optical bipolar jet of a comparable position angle (Neckel & Staude 1995). Moreover, the orientation of the K-band (Neckel & Staude 1995) emission of the field as well as the 2MASS IR emission have position angles that are similar to that of the radio emission, suggesting that they trace the bipolar jet. Felli et al. (1998) observed the field at 8.4 and 15 GHz and detected weak radio emission at the position marked C in the upper panel of Figure 13. Purser et al. (2016), on the other hand, detected both C and W at 5.5 and 9.0 GHz where they established that C is thermal while W is non-thermal, implying that the MeerKAT emission is due to a jet-lobe system. Given the proximity of C (Felli et al. 1998, Purser et al. 2016) to the location of the IR source (black asterisk), it appears to be the core that drives the jet. Source W, however, is offset from the IR source confirming that it is a jet lobe as classified by Purser et al. (2016). Additionally, the SED of the object, shown on the lower panel of Figure 13, manifests a decline in flux from 1.3 GHz to 5.5 GHz followed by a rise to 15 GHz, implying the presence of non-thermal emission at 1.3 GHz.

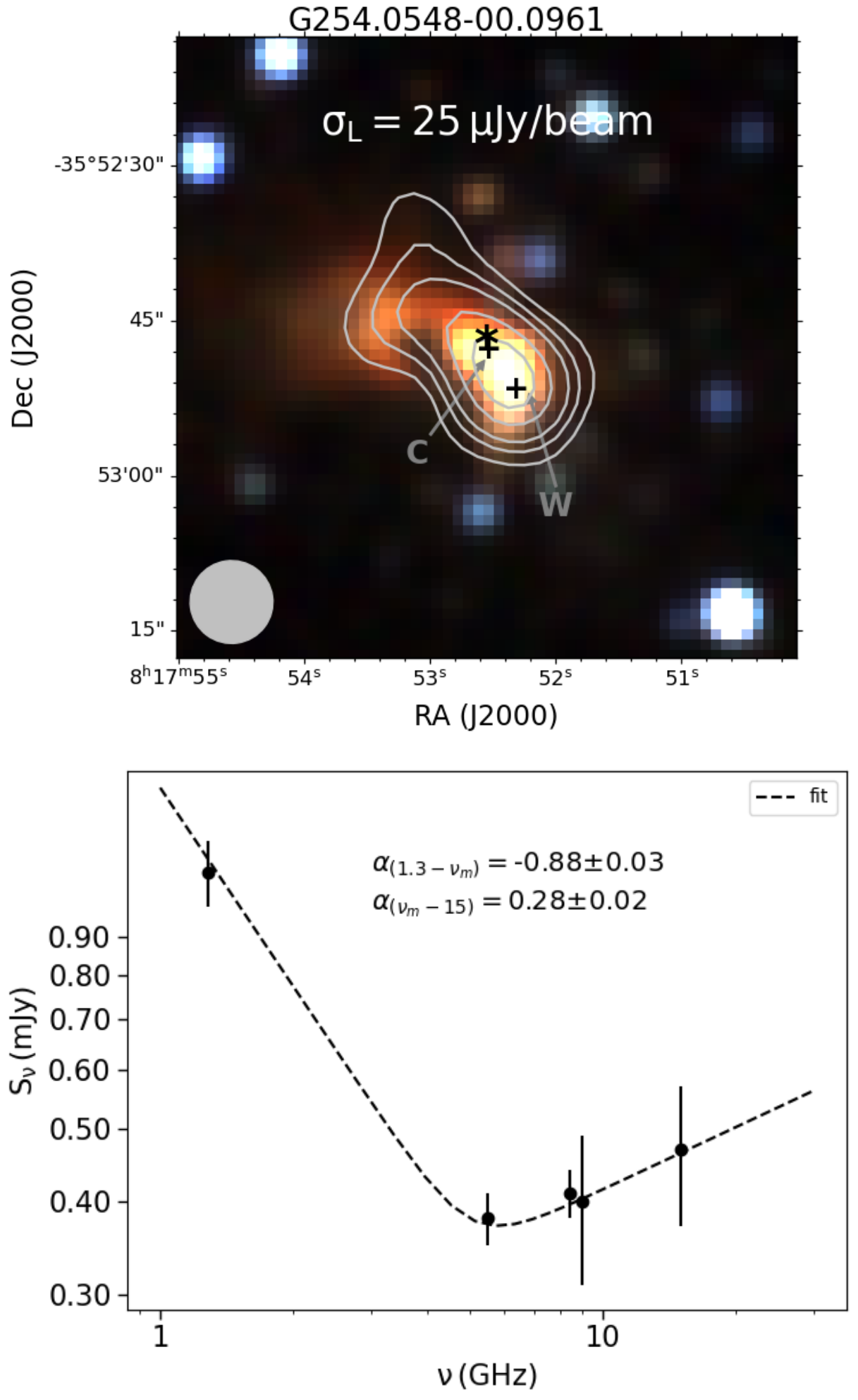

**Figure 13.** Top: 2MASS three-colour map of G254.0548-00.0961's field with 1.3 GHz emission contours over-plotted at levels 3,5,7,11 and 15$\sigma$. The plus signs represent the location of radio sources detected by Felli et al. (1998) and Purser et al. (2016). The location of the IR source in the field is also marked by the asterisk. The grey ellipse shows the synthesised beam of the MeerKAT observation. Bottom: SED of G254.0548-00.0961.

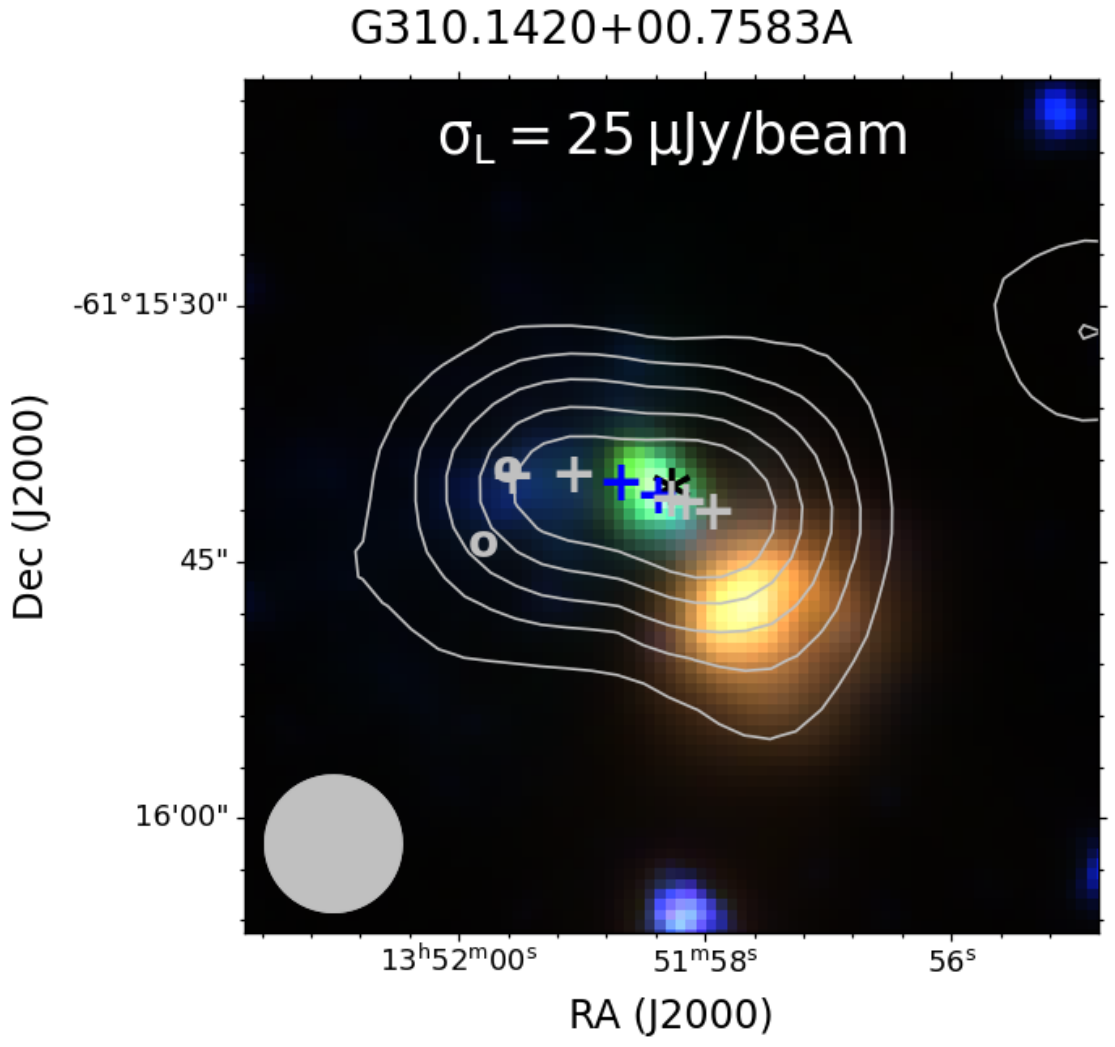

**Figure 14.** The GLIMPSE three-colour map of the field at 3.6 $\mu$m, 5.8 $\mu$m and 8.0 $\mu$m, overlaid with contours of 1.3 GHz emission at 3, 7, 15, 30 and 65$\sigma$. The blue and grey plus(+) signs represent the locations of thermal and non-thermal radio sources respectively (Purser et al. 2016). The IR source in the field is denoted by the asterisk while the grey circular symbols represent the locations of 2.12 $\mu$m knots detected in Caratti o Garatti et al. (2015). The grey ellipse represents the synthesised beam of the MeerKAT observation.

**4.2.0.6 G310.1420+00.7583A:** This source displays a single extended emission of flux density 11.0±0.7 mJy at 1.3 GHz. Its radio emission is aligned in an E-W direction. The studies by Cyganowski et al. (2008), Caratti o Garatti et al. (2015) and Purser et al. (2016) all point to the presence of a jet in its field. Cyganowski et al. (2008) detected an extended green object (EGO), Caratti o Garatti et al. (2015) observed two near-IR knots at 2.12 $\mu m$ and Purser et al. (2016) discovered seven radio sources in the field. In their 5.5, 9.0, 17.0 and 22.8 GHz maps, Purser et al. (2016) detected two thermal radio sources marked with blue plus (+) signs, and five other non-thermal radio sources marked with grey plus signs (see Figure 14). All the radio sources are aligned with the emission at 1.3 GHz. The location of one of the thermal sources is coincident with the position of the IR source (asterisk) suggesting that it is the jet driver. The 2.12 $\mu m$ knots (Caratti o Garatti et al. 2015), marked by O on Figure 14, also show alignment with the jet, especially the one on the northern part, suggesting that they are associated with the jet. While the jet's emission is comprised of radiation from both thermal and non-thermal components (Purser et al. 2016), its spectral index, calculated using the MeerKAT and Purser et al. (2016) observations, is -0.14±0.03, suggesting that it is a non-thermal emitter.

**4.2.0.7 G343.1261-00.0623:** Three radio sources denoted by C, NN and SS on Figure 15 were detected in the field of G343.1261-00.0623. Their flux densities at 1.3 GHz are 33.2±2.5 mJy, 1.04±0.09 mJy and 0.88±0.05 mJy respectively. The three radio emitters are aligned at a position angle of $\theta_{PA}$ ~ 161±2°. Purser et al. (2016) detected seven radio components of C at 17 and 22 GHz. One of the components was classified as thermal and the rest as non-thermal, depicting a morphology of a thermal protostellar core and six non-thermal lobes. Earlier, Rodríguez et al. (2008)'s high-resolution observation detected twelve components of C at 8.3 GHz, classifying a component as a jet-core and the rest as lobes. All their lobes were classified as non-thermal except one. NN and SS have spectral indices $\alpha$ <-0.34 and <-0.39 respectively. The upper limits on the indices of NN and SS were estimated from the 3$\sigma$ noise of 5.5 GHz observation in Purser et al. (2016). It is not clear if these two objects are related to the jet, C, however, their spectral indices, consistency with the precession model in Rodríguez et al. (2008) and symmetry about the thermal core suggests that they are non-thermal lobes of the jet. NN and SS, are separated by an angular displacement $\theta \simeq 86''$, equivalent to a physical separation of ~1.2±0.3 pc on the plane of the sky, assuming that the jet is located at a distance $d$ =2.9±0.6 kpc (Rodríguez et al. 2008) from the earth. The spectral index of C, estimated from MeerKAT, Rodríguez et al. (2008) and Purser et al. (2016) fluxes, is -0.17±0.09, confirming that it emits non-thermal radiation.

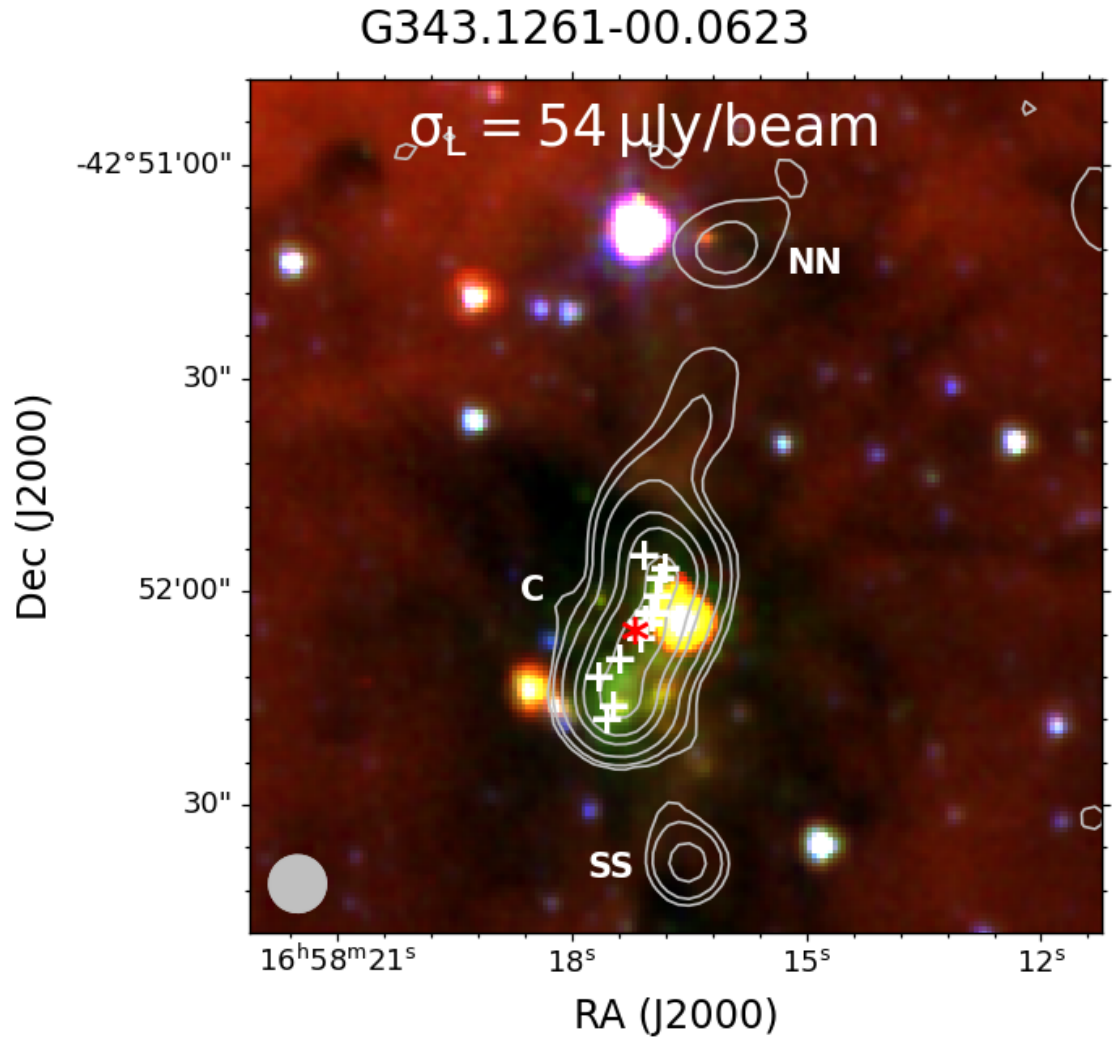

**Figure 15.** GLIMPSE's three colour map of the field at 3.6 $\mu$m, 5.8 $\mu$m and 8.0 $\mu$m with an overlay of contours of 1.3 GHz emission at levels 3,5,11,30, 50 and 100$\sigma$. The asterisk and plus(+) signs represent the locations (Rodríguez et al. 2008) of the thermal core and its non-thermal radio lobes respectively. The grey ellipse shows the synthesised beam of the MeerKAT observation.

**4.2.0.8 G345.4938+01.4677:** Three radio sources of flux densities 51.8±0.8 mJy, 31.9±1.1 mJy and 29.9±0.9 mJy, denoted as C, OE and OW, respectively on Figure 16 were detected in the field of G345.4938+01.4677 at 1.3 GHz. The three sources are aligned at a position angle, PA= 103±5°. Both Guzmán et al. (2010) and Urquhart et al. (2007) detected five radio sources in the field, the core and four lobes. C hosts the core and two inner lobes denoted as IW and IE in Guzmán et al. (2010). However, OW and OE were not detected by Purser et al. (2016). A plot of the SED of C using fluxes from MeerKAT, Guzmán et al. (2010), Urquhart et al. (2007) and Purser et al. (2016) shows that its flux at 1.3 GHz is higher as can be seen on the lower panel of Figure 16, perhaps signifying the presence of non-thermal emission at lower frequencies. The spectral indices of lobes OE and OW were estimated to be ≃ -0.48±0.46 and -1.45±0.41 respectively, suggesting that they are non-thermal.

Both Guzmán et al. (2010) and Purser et al. (2016) classified the object as a jet composed of a thermal core of spectral index $\alpha$~0.8 and associated lobes. More recently, Guzmán et al. (2016) classified the driving source as a hyper-compact HII-region (HCHII) that harbours a disk-like structure (Guzmán et al. 2020). Infrared observation of the field also manifested the presence of a jet (Guzmán et al. 2010). The position of OE's peak emission in Urquhart et al. (2007) is offset from that in the MeerKAT observation, perhaps signifying precession.

### 4.3 Spectral indices of all the sources

Most of the MYSOs in the MeerKAT GPS were observed at $\nu \simeq$ 5 GHz by Becker et al. (1994), Urquhart et al. (2007) and Purcell et al. (2013), and some of them at $\nu$ = 8.6 GHz by Urquhart et al. (2007). A few of the sources were also observed by Purser et al.

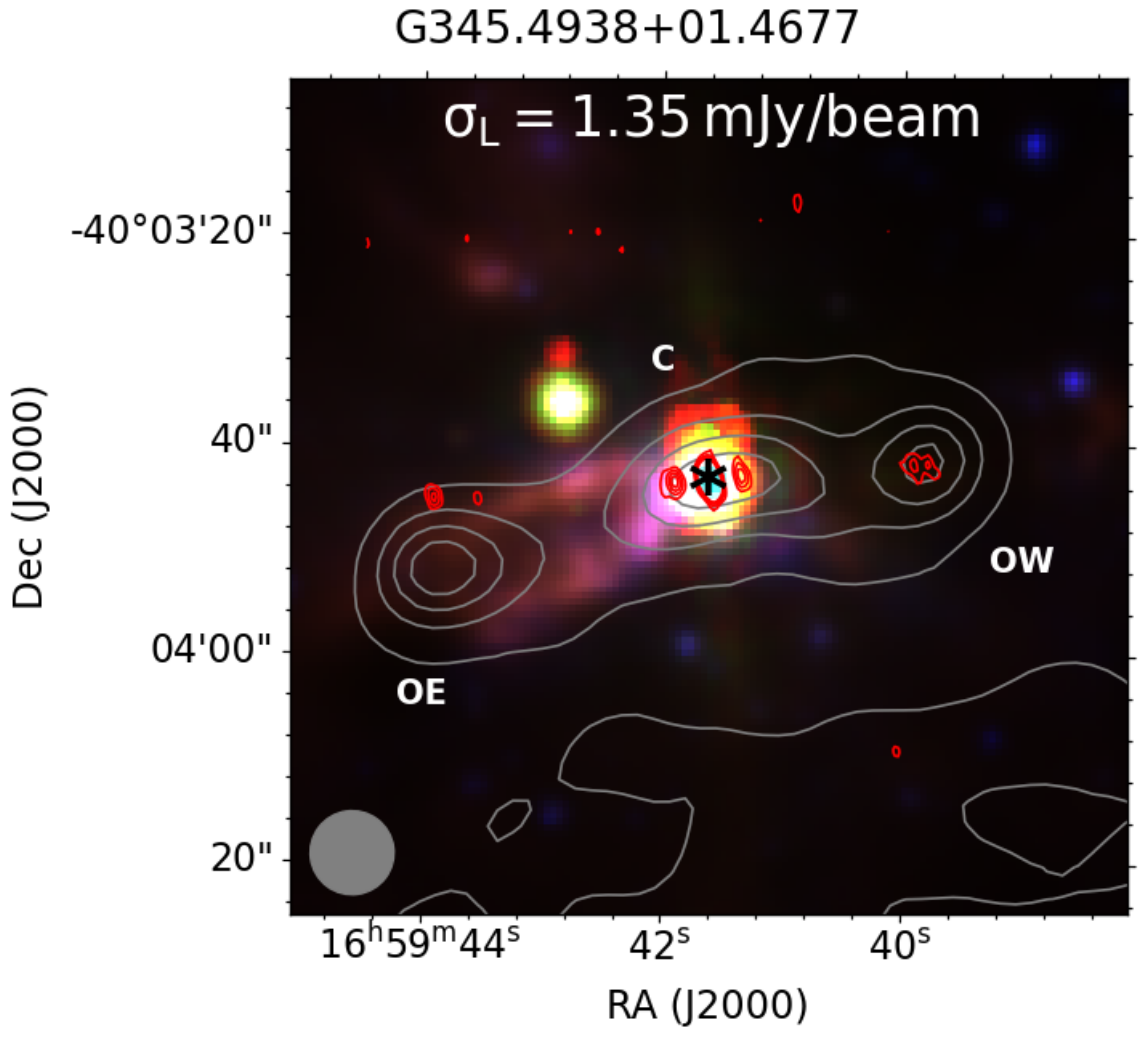

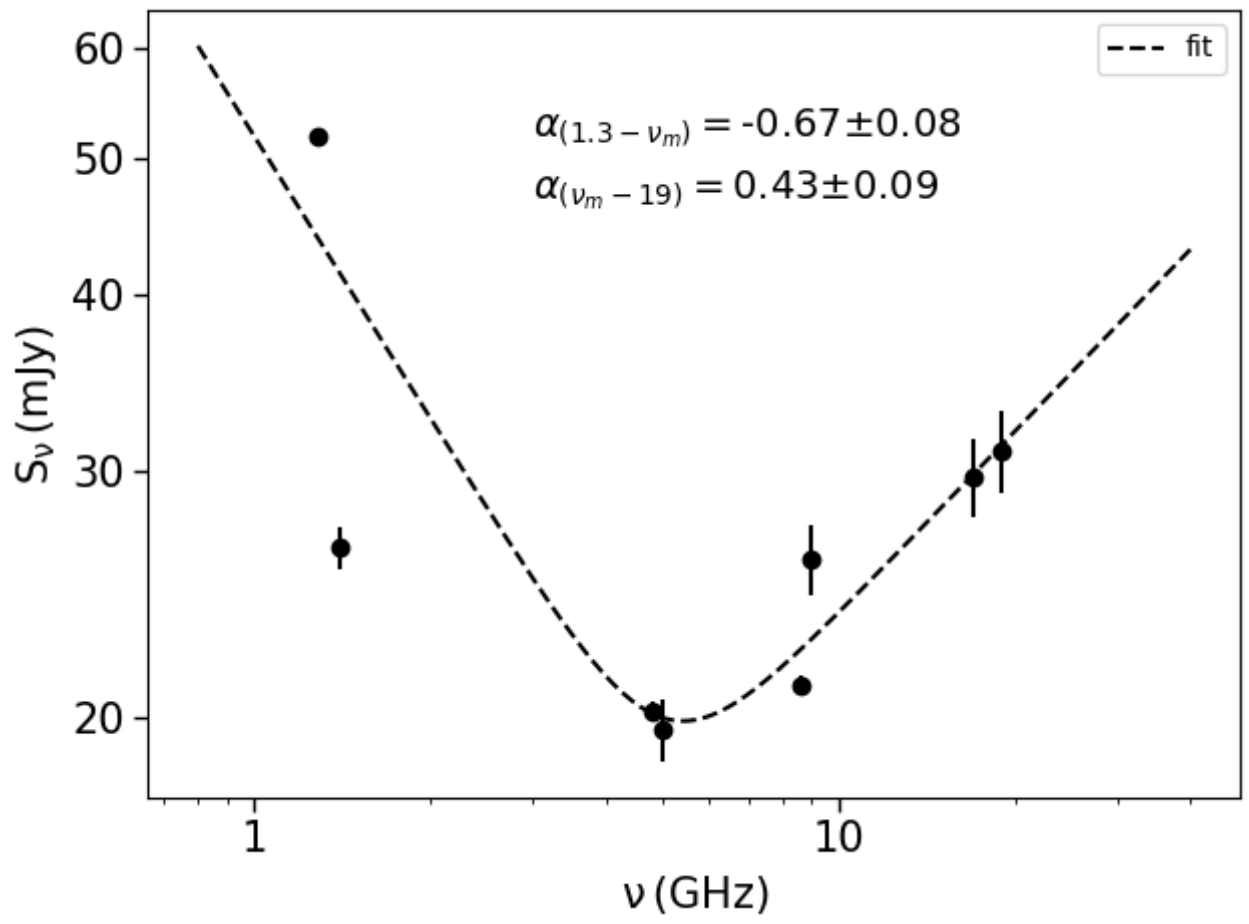

**Figure 16.** Top: GLIMPSE's three colour map of the field at 3.6 $\mu$m, 5.8 $\mu$m and 8.0 $\mu$m with overlay of grey and red contours of 1.3 GHz and 4.8 GHz (Urquhart et al. 2007) emission at levels 5, 7, 9, 13 & 15$\sigma$ and 3, 5, 7 & 11$\sigma$ respectively. The asterisk sign represents the locations of the thermal core (Guzmán et al. 2010, Purser et al. 2016). The grey ellipse shows the synthesised beam of the MeerKAT observation. Bottom: SED of G345.4938+01.4677.

(2016) at 5.5 GHz, 9.0 GHz, 17.0 GHz and 22.8 GHz (see Table A1). These observations, and any other radio fluxes that are available in the literature, were used to calculate the spectral indices of the jets. The indices were calculated from the slopes of their flux versus frequency plots, in logarithmic scale, if they were detected in at least two observations of different frequencies, otherwise a limit, calculated from 3$\sigma$ of the rms noise of their fields, was set on their indices. If a protostar's field was observed in two or more observations, at the same frequency, without detection, then the rms of the most sensitive observation was used in setting a limit on the protostar's flux and spectral index.

The derived spectral indices range from $\alpha_\nu$ =-0.86 to 1.35, showing that the sample consists of non-thermal lobes ($\alpha_\nu$ <-0.1), thermal cores of protostellar jets ($\alpha_\nu$ ≃0.6) and possibly HII regions. The histogram shown in Figure 17, displays two peaks, at $\alpha \simeq -0.53$, and $\alpha \simeq 0.37$, which are indicative of non-thermal and thermal emitters, respectively. Most of the thermal emitters have spectral indices in the range $0.02 \leq \alpha_\nu \leq 0.70$, typical of ionized protostellar jets. Whereas most of the indices are calculated from observations of different epochs and may be uncertain due to variability in jets, their spectral energy distributions (SEDs) show consistency with the plotted fluxes e.g the SEDs of G310.0135+00.3892 and G339.6221-00.120 shown in Figure 18 that were generated with fluxes from this work, Purcell et al. (2013) and Purser et al. (2016). Similarly, the different observations may be sensitive to different spatial scales, as a result, some of the fluxes used in calculating the indices may be due to emission of dissimilar features of the objects. The relationship between the indices and the morphology of the sources was explored by plotting their indices against the ratio of flux-to-peak intensity (see the right panel of Figure 17). This is because a large flux-to-peak intensity ratio is an indicator of how extended a source is (Kavak et al. 2021). The sources associated with negative spectral indices show no trend with size whereas the positive spectral indices sources do appear to be preferentially compact.

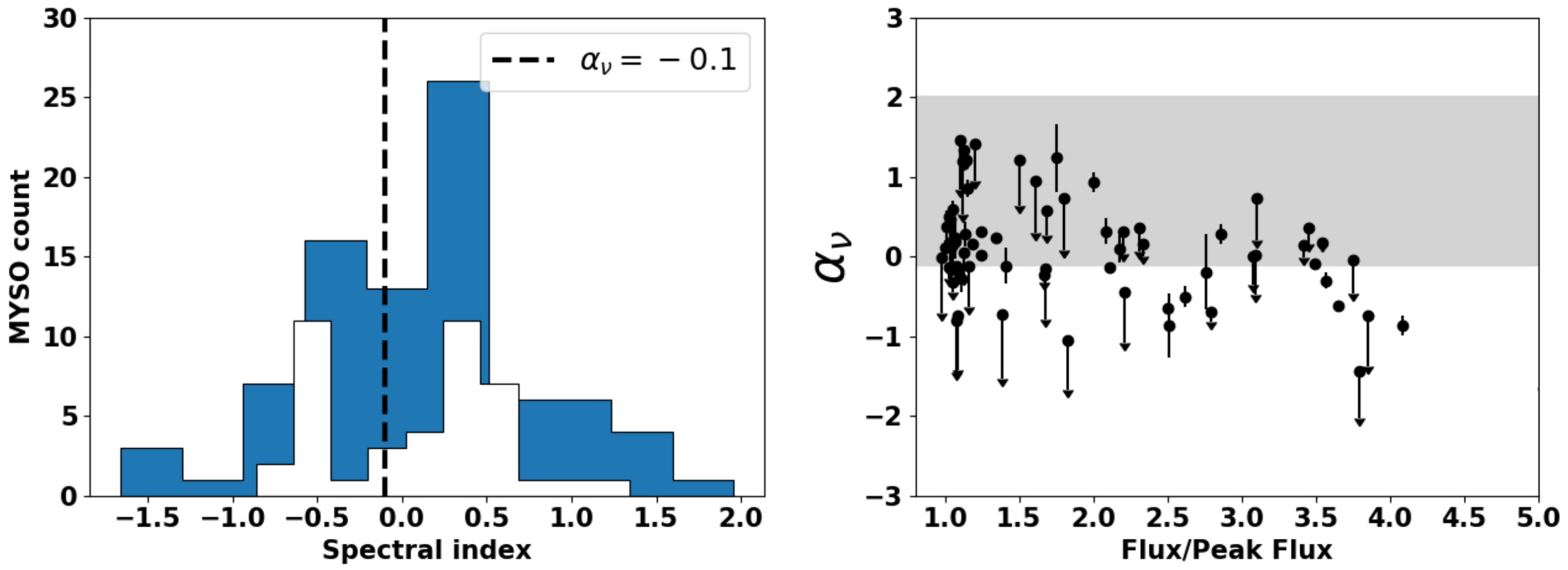

**Figure 17.** Left: Spectral index distribution for the sample of MYSOs in the MeerKAT GPS, calculated using Becker et al. (1994), Urquhart et al. (2007), Purcell et al. (2013) and Purser et al. (2016). The blue histogram is for all the sources, including upper and lower limits, while the white one is for sources with definite indices. The dashed line shows the boundary between the indices of thermal and non-thermal radiation. Right: Spectral index versus flux-to-peak intensity ratio, calculated from MeerKAT data, for the sources with definite indices. The upper limits of some of the sources are also shown. The region that is shown in grey represents the spectral index regime for thermal-free emission.

In some objects, the Spectral Energy Distribution (SED) exhibited a decline with frequency followed by a rise, and vice versa. To model their SEDs, Reynolds (1986)'s model for conical, ionised winds and Dougherty et al. (2003)'s model for synchrotron emission were combined to generate a broken power law function represented by equation 2. In the equation, $\nu_\mathrm{m}$ represents the turnover frequency of the SED, the point at which the two models intersect. The parameters $\alpha_1$ and $\alpha_2$ are the spectral indices below and above the $\nu_m$, respectively. $\Delta$ is a parameter that governs the smoothness of the transition between the two slopes at the point of intersection (see the lower panels of Figures 10, 11, 13, and 16). Least square fits on the data points were performed by first assigning initial guesses, using the negative indices, positive indices, and minimum fluxes as starting points.

$$S(\nu) \propto \left(\frac{\nu}{\nu_m}\right)^{\alpha_1} \left\{\frac{1}{2}\left[1+\left(\frac{\nu}{\nu_m}\right)^{1/\Delta}\right]\right\}^{(\alpha_1-\alpha_2)\Delta} \quad (2)$$

To determine the proportion of jets associated with non-thermal emission, the spectral indices of the MYSOs below and above $\nu_m$ were compared. The left panel of Figure 19 shows that 42% of the jets have negative indices below $\nu_m$, and positive ones at higher frequencies, implying the presence of non-thermal emission. Two of the sources, on the other hand, exhibit negative indices at all frequencies, such as G339.6221-00.1209. Overall, 52% of the sources with fluxes in more than two frequencies display evidence of non-thermal emission. Even the SED of the unresolved source G318.9480-00.1969A at 1.3 GHz, shown on the right panel of Figure 19, suggests that it emits non-thermal emission and is a potential jet driver. This finding is consistent with the study by Lee et al. (2001) and De Buizer (2003), who detected $H_2$ knots associated with the MYSO.

### 4.4 Radio Luminosity

Empirical studies show that ionizing radiation of low-mass protostars is insufficient to generate the observed radio luminosities in the cores of the protostellar jets. Additionally, the cores of high-mass protostellar jets exhibit radio luminosities lower than what is expected from photo-ionization (Anglada et al. 2018). The discrepancy in radio-luminosities of protostellar jets and HII-regions has been widely used to distinguish jets from ultra-compact HII-regions, which are typically brighter than the jets by at least two orders of magnitude (Kalcheva et al. 2018). For instance, UCHII-regions have radio-luminosities in the range of $10^2$ – $10^5$ mJykpc$^2$ (Kalcheva et al. 2018) while protostellar jets have luminosities around ~ 0.1 - 30 mJykpc$^2$ (Purser et al. 2016, Anglada et al. 2018, Obonyo et al. 2019).

The MYSOs in the SARAO MeerKAT GPS survey exhibit a similar trend, having lower radio-luminosities compared to those of UCHIIs, and showing a correlation with the luminosities of low-mass protostars as illustrated in Figure 20. However, their radio-luminosities are generally higher than those reported in previous studies, which were conducted at higher frequencies and resolutions. Thermal emitters in this sample have an average radio-luminosity of 5 mJykpc$^2$ at 1.3 GHz, comparable to the average radio luminosity of 8.3 GHz reported by Anglada et al. (2018) and slightly higher than the 2.5 mJykpc$^2$ estimated by Obonyo et al. (2019) at 1.5 GHz. Purser et al. (2016) estimated the luminosity at 9.0 GHz, finding an average value of 10 mJykpc$^2$, with some MYSOs reaching as high as 30 mJykpc$^2$. Non-thermal emitters in the MeerKAT sample, however, display an average radio-luminosity approximately an order of magnitude higher than the brightest MYSO in Obonyo et al. (2019)'s study at a similar frequency band, i.e, ~100 mJykpc$^2$.

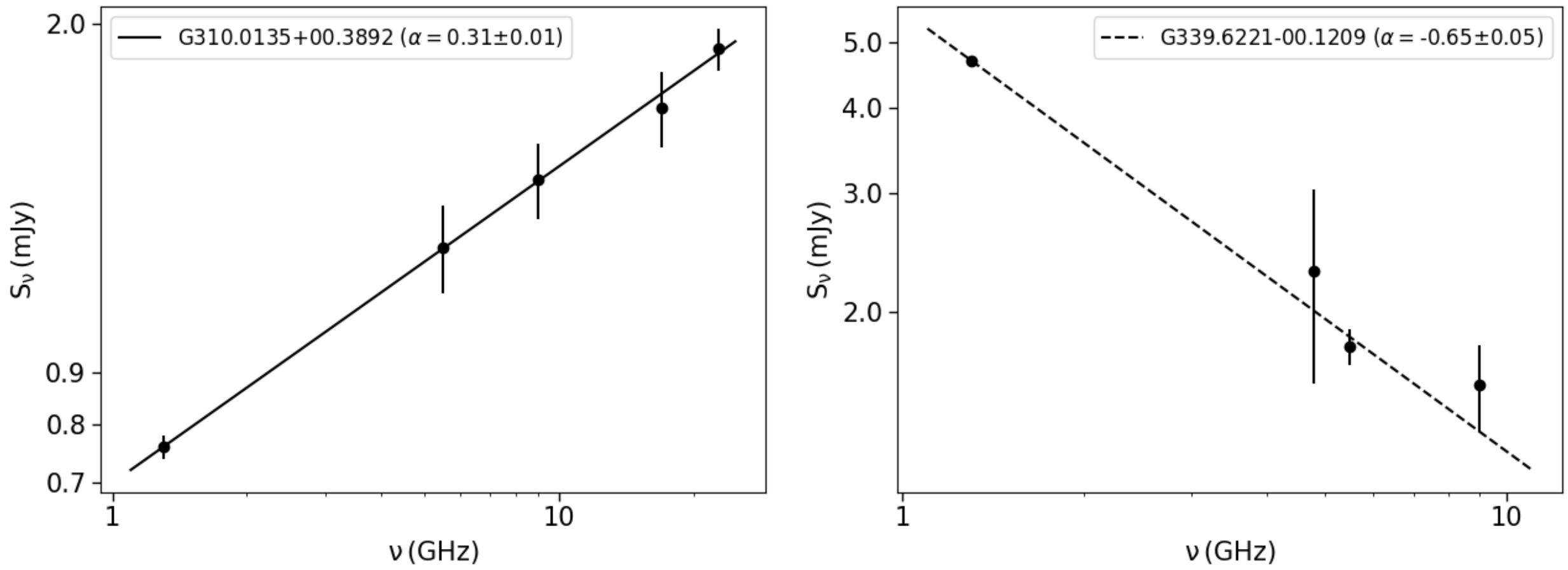

**Figure 18.** The spectral energy distributions of G310.0135+00.3892 (left) and G339.6221-00.1209 (right) illustrate how the spectral indices of thermal (left) and non-thermal (right) sources were estimated from the slopes of the $S_\nu$ vs $\nu$ plots. The flux densities were obtained from MeerKAT observation, as well as from Purcell et al. (2013), and Purser et al. (2016).

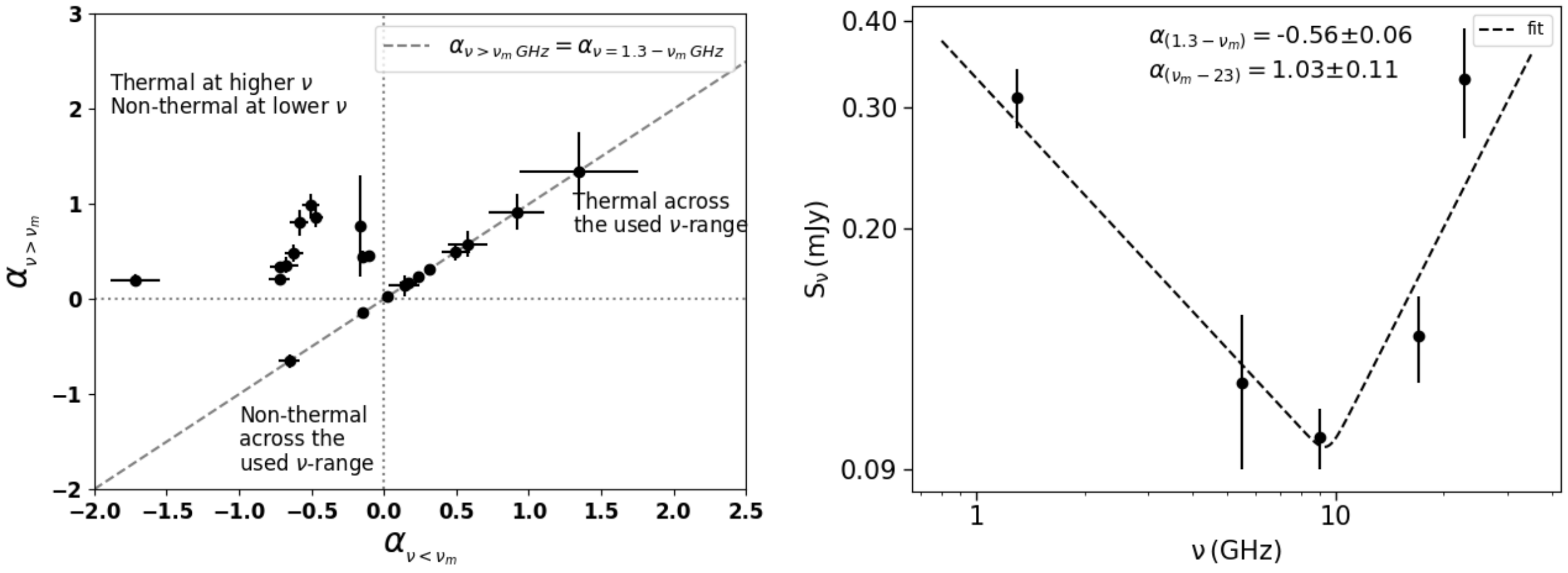

**Figure 19.** Left: A plot of $\alpha_{\nu<\nu_{\rm m}}$ vs $\alpha_{\nu>\nu_{\rm m}}$ for the MYSOs in the MeerKAT GPS. The dashed line represents points where the two indices are equal, indicating cases where the spectral index is constant across all frequencies. Right: SED of G318.9480-00.1969A showing the model (dashed line) used to estimate its spectral indices below and above $\nu_{\rm m}$.

The difference in the radio-luminosities between the MeerKAT and the JVLA[3] observation of similar objects (Obonyo et al. 2019) can be attributed to their different resolutions. For instance, the angular sizes of the cores in Obonyo et al. (2019) are typically $\leq 3''$ whereas in the MeerKAT sample, the minimum, average and maximum major axes are 2.5″, 12.4″ and 30.1″ respectively, with only one source having a major axis $\theta \leq 3''$. In addition, all the lobes in their study were located $\lesssim 5''$ from the cores, suggesting that a significant portion of the emission in the MeerKAT sample may originate from both thermal cores and non-thermal lobes, as seen in G035.1979-00.7427 and G310.1420+00.7583A (see Figures 11 and 14).

Furthermore, the majority of the MYSOs associated with negative spectral indices have higher radio-luminosities compared to those with positive indices (see Figure 20), perhaps an indication of a mixture of fluxes from their unresolved thermal and non-thermal components. Kavak et al. (2021) also noted that sources with negative spectral indices tend to be more luminous at radio wavelengths even at higher frequencies, such as 6 GHz. Indeed, a recent study by Carrasco-González et al. (2021) that zoomed into the collimation zone of the Cep A HW2 radio jet suggests that at MeerKAT resolution, both lobe and core emission are present in some of the radio sources. Consequently, most sources have luminosities that steepen the gradient of bolometric-versus-radio-luminosity plots for protostars of all masses, possibly due to the presence of emission from unresolved lobes. This phenomenon appears to be minimal in higher-frequency observations, as non-thermal emission is weaker at those frequencies. Moreover, higher resolution observations can resolve the lobes, thus separating the thermal cores from the lobes.

[3] Jansky Very Large Array

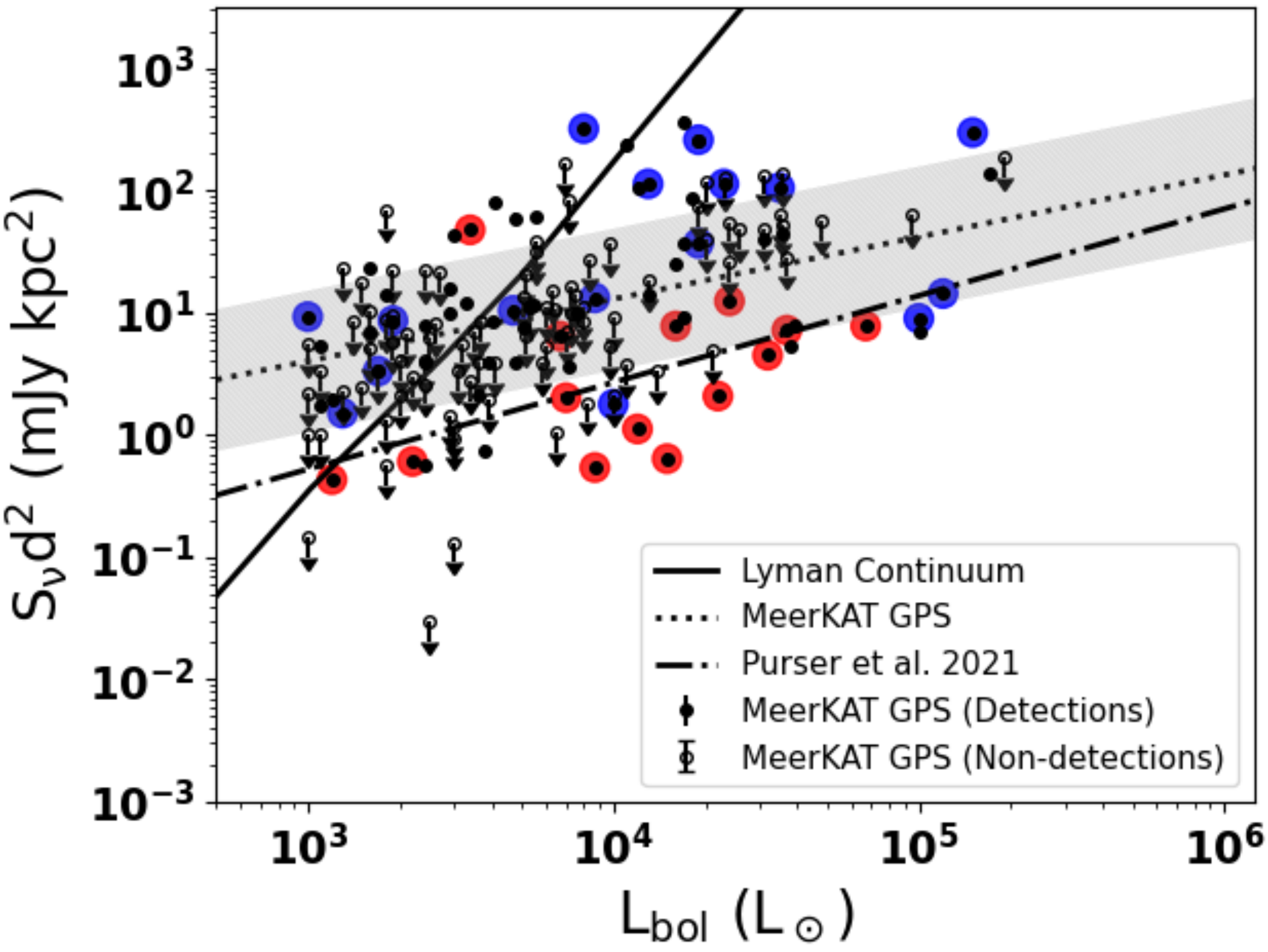

**Figure 20.** Correlation between radio and bolometric luminosity at 1.3 GHz. The dotted line is a least-squares fit to MeerKAT data and the grey area indicates the residual standard deviation of the fit. The dot-dashed line is a fit to the high-mass objects from Purser et al. (2016). The solid line corresponds to the expected radio luminosity of an optically thin region photoionized by the Lyman continuum of a star. The red and blue points represent thermal and non-thermal emitters respectively.

### 4.5 Evolution of ionized jets

Chambers et al. (2009) used infrared observations at 4.5 $\mu$m and 24 $\mu$m to determine the evolutionary stages of massive protostellar cores within Infrared Dark Clouds (IRDCs). They classified the youngest cores in their sample as quiescent if they were undetected at both wavelengths, while those exhibiting extended 4.5 $\mu$m emission and detected at 24 $\mu$m were considered active. Sources with extended 4.5 $\mu$m emission, known as Extended Green Objects (EGOs; Cyganowski et al. 2008), are associated with outflow shocks. These emission likely arise from fundamental ro-vibrational CO lines around 4.7 microns, whereas the 24 $\mu$m emission may result from heated dust in the core caused by the conversion of gravitational energy of the infalling materials into thermal energy (Chambers et al. 2009). Moreover, Guzmán et al. (2015) observed that deeply embedded protostars, which do not exhibit star formation activities at 24 $\mu$m, may be detectable at 70 $\mu$m. Consequently, the ratio of flux densities at 70 $\mu$m to 24 $\mu$m can serve as an indicator of the earliest stages of star formation.

Studies by Purser et al. (2016), Purser et al. (2021), Obonyo et al. (2019), and this work indicate that non-thermal emitters constitute approximately 40-50% of the ionized jets, suggesting that thermal and non-thermal jets may represent different evolutionary stages. To investigate this further, we computed the ratio of 70 $\mu$m (Elia et al. 2017) to 24 $\mu$m (Gutermuth & Heyer 2015) fluxes and plotted them against their bolometric luminosities (see Figure 21). The two populations do not show clear separation, indicating that both thermal and non-thermal emitters are present at all phases of evolution. However, their distribution suggests that thermal and non-thermal emission may preferentially arise at different stages. Theoretically, early MYSO phases may be characterized by swollen, cool, convective envelopes and jet-driving. As the MYSO evolves, the central star contracts, heats up, and radiatively drive a disc wind (Hoare 2006, Gibb & Hoare 2007) in addition to the jet.

## 5 CONCLUSIONS

The SARAO MeerKAT Galactic Plane Survey has provided the largest and a highly sensitive observation of massive protostars at 1.3 GHz to date. A total of seventy-one protostellar jets were detected, while ninety-one showed non-detections, and the remaining protostellar fields were heavily affected by nearby bright HII regions. The majority of the detected MYSOs exhibited weak radio emission with integrated fluxes $F_{int} \leq 1.5$ mJy. Analysis of their spectral indices indicates that approximately 52% of the sample emits non-thermal radio emission. Moreover, examining the SED profiles of the MYSOs with fluxes in more than two frequencies reveals that at least 42% of the objects manifest a turnover between non-thermal and thermal parts of the SEDs. Additionally, the radio-luminosities of some of the objects indicate a combination of thermal and non-thermal radiation. These findings agree with the outflow model of massive protostars described in Carrasco-González et al. (2021), with the sources previously observed at higher frequencies showing multiple non-thermal lobes, most of which are enveloped by the extended emission in the MeerKAT observation.

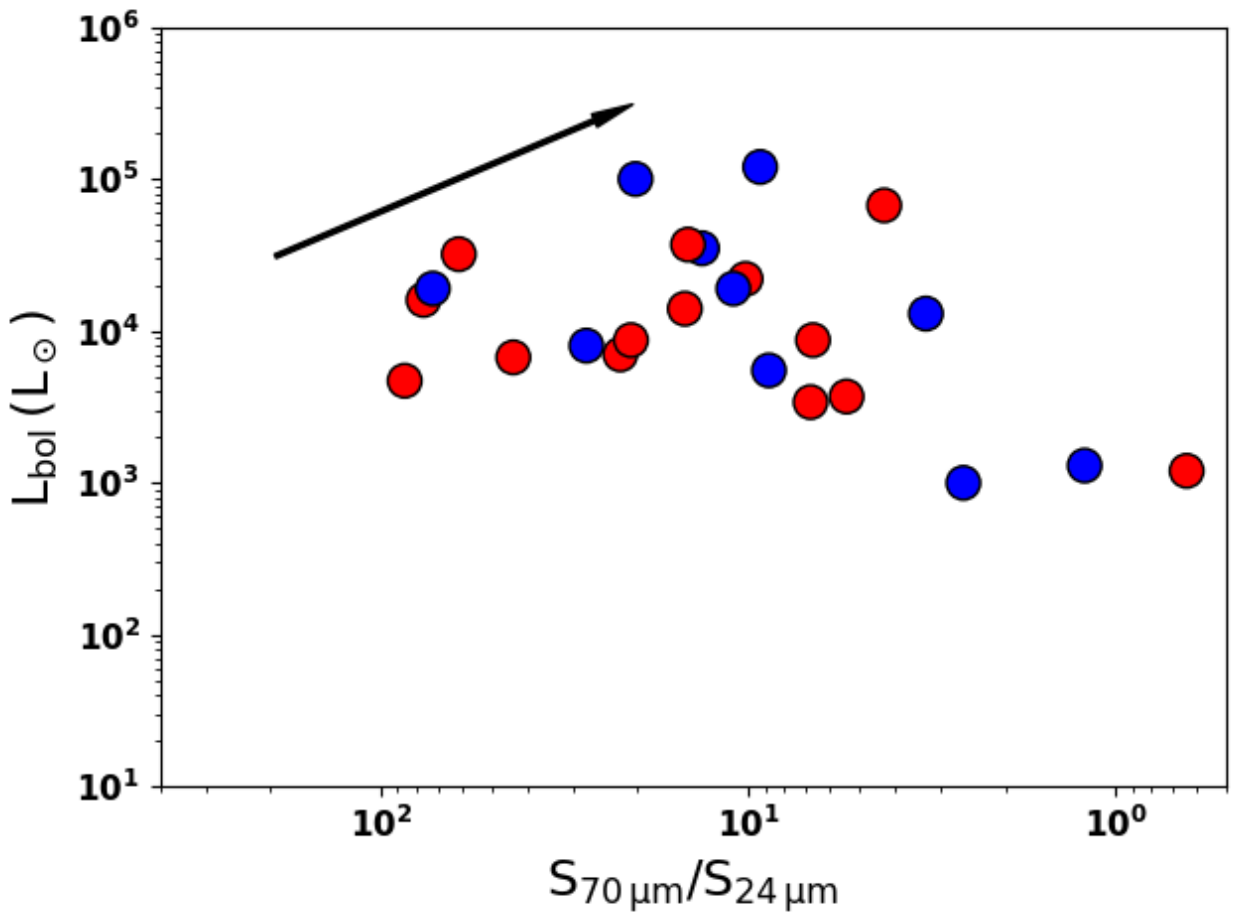

**Figure 21.** A plot of bolometric luminosity vs mid-infrared, 70 $\mu$m/24 $\mu$m, flux ratio for some of the objects in the sample. The blue and red points represent non-thermal and thermal emitters respectively.

## ACKNOWLEDGEMENTS

The MeerKAT telescope is operated by the South African Radio Astronomy Observatory, which is a facility of the National Research Foundation, an agency of the Department of Science and Innovation. WOO acknowledges research support by the DARA project.

## DATA AVAILABILITY

Data Availability Statement: The data underlying this paper are available at the SARAO Data Archive: (`https://archive.sarao.ac.za/`) under the project code SSV-20180721-FC-01.

## APPENDIX A: TABLES

Table A1: Massive protostars in the MeerKAT GPS, their positions at 1.3 GHz, integrated fluxes, sizes, offsets from IR sources, and spectral indices. The references used in estimating or setting limits on the spectral indices are given in the last column.

| MYSO | Position | | flux (mJy) | size | | PA (°) | IR offset | $\alpha_\nu$ | Refs |
|---|---|---|---|---|---|---|---|---|---|
| | RA(J2000) | DEC(J2000) | | $\theta_{maj}('')$ | $\theta_{min}('')$ | | ($''$) | | |
| G010.8856+00.1221 | $18^h09'08.04''$ | -19°27′24.4″ | 1.6±0.13 | 13.0±1.0 | 9.0±1.0 | 32±10 | 0.94 | -0.19±0.47 | 6 |
| G011.5001-01.4857 | $18^h16'22.43''$ | -19°41′25.3″ | 0.21±0.07 | 10.2±0.4 | 5.7±0.3 | 21±5 | 2.26 | 0.19±0.27 | 2 |
| G012.0260-00.0317 | – | – | <0.51 | – | – | – | – | >1.96 | 3 |
| G012.7879-00.1786 | $18^h14'06.55''$ | -17°56′05.6″ | 10.6±1.2 | 15.2±0.4 | 14.4±0.4 | 52±22 | 1.34 | <-0.80 | 5 |
| G012.8909+00.4938A | – | – | <0.18 | – | – | – | – | – | – |
| G012.8909+00.4938C | $18^h11'51.44''$ | -17°31′30.2″ | 0.35±0.04 | – | – | – | 0.10 | 0.5±0.06 | 3 |
| G012.9090-00.2607 | $18^h14'39.49''$ | -17°52′01.1″ | 0.78±0.14 | – | – | – | 0.47 | 0.24±0.02 | 4 |
| G013.3310-00.0407 | – | – | <0.42 | – | – | – | – | – | – |
| G013.6562-00.5997 | – | – | <0.20 | – | – | – | – | – | – |
| G015.0939+00.1913 | $18^h17'20.85''$ | -15°43′46.3″ | 2.73±0.06 | 14.7±0.4 | 12.2±0.3 | 31±5 | 0.49 | <-1.44 | 5 |
| G016.7122+01.3119 | – | – | <0.30 | – | – | – | – | – | – |
| G016.9261+00.2854 | – | – | <0.40 | – | – | – | – | – | – |
| G017.4507+00.8118A | $18^h19'41.96''$ | -13°21′37.8″ | 1.22±0.02 | 18.6±0.4 | 14.2±0.3 | 4.0±0.5 | 2.67 | <-0.74 | 6 |
| G017.6380+00.1566 | $18^h22'26.23''$ | -13°30′10.4″ | 1.86±0.29 | 15.1±2.7 | 7.3±1.9 | 130±10 | 2.59 | 0.03 ± 0.11 | 5,7 |
| G017.9642+00.0798B | – | – | <0.21 | – | – | – | – | – | – |
| G017.9642+00.0798C | – | – | <0.21 | – | – | – | – | – | – |
| G017.9789+00.2335A | $18^h22'49.18''$ | -13°10′01.6″ | 0.19±0.03 | – | – | – | 0.59 | <1.20 | 5 |
| G018.6608+00.0372A | $18^h24'50.27''$ | -12°39′20.0″ | 0.21±0.03 | – | – | – | 2.44 | <0.94 | 5 |
| G019.8817-00.5347 | $18^h29'14.45''$ | -11°50′23.3″ | 0.96±0.09 | 10.5±0.1 | 6.2±0.1 | 93.5±0.9 | 3.39 | 0.32±0.17 | 3 |
| G019.9224-00.2577 | – | – | <0.18 | – | – | – | – | – | – |
| G020.5703-00.8017 | – | – | <0.60 | – | – | – | – | – | – |
| G021.5624-00.0329 | – | – | <0.60 | – | – | – | – | – | – |
| G023.6566-00.1273 | – | – | <0.21 | – | – | – | – | – | – |
| G028.1467-00.0040A | – | – | <0.71 | – | – | – | – | – | – |
| G029.4375-00.1741 | – | – | <0.57 | – | – | – | – | – | – |
| G029.5904-00.6144 | – | – | <0.42 | – | – | – | – | > 0.97 | 15 |
| G032.0451+00.0589 | $18^h49'36.59''$ | -00°45′45.5″ | 0.52±0.08 | – | – | – | 0.46 | 0.39±0.22 | 15 |
| G032.0518-00.0902 | $18^h50'09.40''$ | -00°49′29.8″ | 2.70±0.25 | – | – | – | 2.36 | 0.38±0.19 | 5 |
| G032.8205-00.3300 | – | – | <0.42 | – | – | – | – | – | – |
| G032.9957+00.0415A | – | – | <1.50 | – | – | – | – | – | – |
| G033.3891+00.1989 | – | – | <0.75 | – | – | – | – | – | – |
| G034.0126-00.2832 | – | – | <0.81 | – | – | – | – | – | – |
| G034.0500-00.2977 | $18^h54'32.28''$ | +00°51′34.4″ | 0.68±0.07 | 16.7±0.1 | 12.5±0.2 | 21.5±0.5 | 1.50 | <-0.13 | 5 |
| G034.8211+00.3519 | – | – | <2.10 | – | – | – | – | >-1.25 | 5 |
| G035.1979-00.7427 | $18^h58'13.08''$ | +01°40′35.7″ | 21.8±2.3 | 19.9±2.3 | 7.4±1.2 | 6.3±4.1 | 4.70 | -0.61±0.05 | 6,9 |
| G035.3449+00.3474 | – | – | <1.20 | – | – | – | – | >-0.32 | 5,15 |
| G035.8546+00.2663 | – | – | <1.30 | – | – | – | – | – | – |
| G037.5536+00.2008 | $18^h59'09.97''$ | +04°12′15.9″ | 0.32±0.07 | – | – | – | 0.54 | 0.23±0.25 | 8 |
| G038.3543-00.9519 | – | – | <0.33 | – | – | – | – | – | – |
| G038.9365-00.4592 | – | – | <0.32 | – | – | – | – | – | – |
| G039.3880-00.1421B | – | – | <1.20 | – | – | – | – | >0.78 | 8 |
| G039.5328-00.1969 | $19^h04'13.58''$ | +05°46′54.6″ | 0.18±0.02 | 10.5±1.1 | 4.6±1.1 | 170.1±5.8 | 0.74 | <0.73 | 5 |
| G040.2816-00.2190 | $19^h05'41.20''$ | +06°26′13.1″ | 0.19±0.03 | – | – | – | 1.74 | 0.59±0.11 | 10 |
| G040.2849-00.2378 | – | – | <0.27 | – | – | – | – | – | – |
| G040.5967-00.7188 | – | – | <0.21 | – | – | – | – | – | – |
| G042.0341+00.1905A | – | – | <0.96 | – | – | – | – | – | – |
| G044.2836-00.5249 | – | – | <0.15 | – | – | – | – | – | – |
| G047.9002+00.0671 | – | – | <0.21 | – | – | – | – | – | – |
| G049.0431-01.0787 | $19^h25'22.31''$ | +13°47′16.0″ | 0.93±0.11 | 13.1±1.7 | 5.7±1.3 | 3.0±6.4 | 3.74 | <-0.45 | 6 |
| G049.5993-00.2488 | – | – | <2.40 | – | – | – | – | – | – |
| G050.2213-00.6063 | $19^h25'57.71''$ | +15°03′00.8″ | 1.31±0.03 | 16.7±0.1 | 10.4±0.1 | 167.6±0.5 | 1.99 | <-0.75 | 6 |
| G052.2078+00.6890 | $19^h25'08.50''$ | +17°24′48.3″ | 1.18±0.13 | 12.8±1.6 | 7.0±1.3 | 158.0±8.3 | 1.00 | -0.86±0.40 | 6 |
| G052.9217+00.4142 | – | – | <0.63 | – | – | – | – | – | – |
| G053.1417+00.0705 | $19^h29'17.51''$ | +17°56′23.1″ | 0.15±0.04 | – | – | – | 1.15 | 0.86±0.11 | 10 |
| G053.5343-00.7943 | – | – | <0.39 | – | – | – | – | – | – |
| G053.5671-00.8653 | – | – | <2.70 | – | – | – | – | – | – |
| G056.3694-00.6333 | $19^h38'31.56''$ | +20°25′18.4″ | 0.38±0.04 | 10.3±0.8 | 8.1±1.0 | 176±11 | 1.03 | -0.11±0.22 | 9 |
| G058.7087+00.6607 | $19^h38'36.81''$ | +23°05′43.8″ | 0.62±0.06 | 8.3±1.1 | 4.9±1.1 | 17±11 | 0.41 | <-0.14 | 6 |

Continued on next page

Table A1 – continued from previous page

| | | | | | | | | | | |
|---|---|---|---|---|---|---|---|---|---|---|
| G059.3614-00.2068 | 19$^h$43′17.98″ | +23°14′02.2″ | 1.92±0.02 | 16.0±0.1 | 12.3±0.1 | 1.2±1.1 | 0.85 | -0.87±0.13 | 6,8 |
| G059.4657-00.0457 | – | – | <0.45 | – | – | – | – | – | – |
| G059.7831+00.0648 | 19$^h$43′10.96″ | +23°44′01.4″ | 0.49±0.09 | 30.1±0.9 | 20.0±0.7 | 44±5 | 4.31 | 0.29±0.16 | 11 |
| G059.8329+00.6729 | 19$^h$40′59.25″ | +24°04′44.5″ | 2.08±0.13 | 7.4±0.7 | 7.1±0.7 | 156±109 | 1.04 | <-1.04 | 6 |
| G254.0491-00.5615 | 08$^h$15′57.19″ | -36°08′08.2″ | 0.37±0.06 | 10.6±2.3 | 7.3±2.0 | 30±23 | 1.63 | 0.10±0.16 | 4 |
| G254.0548-00.0961 | 08$^h$17′52.54″ | -35°52′48.7″ | 1.10±0.11 | 15.4±1.8 | 5.4±1.2 | 44.0±4.1 | 1.47 | -0.50±0.13 | 4,12 |
| G259.4634-01.5424 | – | – | <0.06 | – | – | – | – | – | – |
| G260.6877-01.3930 | – | – | <0.03 | – | – | – | – | – | – |
| G263.7434+00.1161 | 08$^h$48′48.59″ | -43°32′26.8″ | 0.88±0.09 | – | – | – | 2.27 | 0.16±0.05 | 4 |
| G263.7759-00.4281 | 08$^h$46′34.88″ | -43°54′30.3″ | 3.08±0.19 | – | – | – | 0.66 | 0.11±0.10 | 4,6 |
| G268.3957-00.4842 | – | – | <0.27 | – | – | – | – | – | – |
| G269.5205-01.2510 | – | – | <0.30 | – | – | – | – | – | – |
| G270.8247-01.1112 | 09$^h$10′30.79″ | -49°41′26.9″ | 0.09±0.03 | – | – | – | 3.06 | <1.33 | 6 |
| G271.2225-01.7712 | 09$^h$09′11.29″ | -50°25′54.7″ | 0.80±0.08 | 13.6±1.4 | 9.5±1.1 | 32±11 | 1.38 | <-0.01 | 6 |
| G279.4100-01.6681 | – | – | <0.05 | – | – | – | – | – | – |
| G281.2206-01.2556 | 10$^h$01′33.65″ | -56°47′16.8″ | 0.08±0.02 | – | – | – | 1.76 | <1.21 | 6 |
| G281.6909-02.0548 | – | – | <0.90 | – | – | – | – | – | – |
| G282.7848-01.2869 | – | – | <0.78 | – | – | – | – | – | – |
| G282.8969-01.2727 | 10$^h$11′31.61″ | -57°47′06.2″ | 0.75±0.05 | 18.2±1.2 | 8.9±0.7 | 118±3.3 | 2.50 | <-0.03 | 6 |
| G284.6942-00.3600 | – | – | <0.75 | – | – | – | – | – | – |
| G287.3716+00.6444 | 10$^h$48′04.71″ | -58°27′01.4″ | 4.22 ±0.38 | – | – | – | 1.26 | – | – |
| G287.8768-01.3618 | 10$^h$44′17.77″ | -60°27′45.6″ | 6.85±0.13 | 17.4±0.3 | 14.6±0.3 | 31.9±4.3 | 1.25 | <-1.66 | 6 |
| G288.9606+00.2645 | – | – | <0.33 | – | – | – | – | – | – |
| G289.0543+00.0223 | 10$^h$57′43.03″ | -59°44′58.6″ | 0.30±0.04 | 10.1±0.1 | 7.9±0.1 | 46±2 | 0.93 | <0.36 | 6 |
| G290.0105-00.8668A | – | – | <1.20 | – | – | – | – | – | – |
| G290.4750-00.3677 | – | – | <0.60 | – | – | – | – | – | – |
| G293.0352-00.7871 | 11$^h$25′33.01″ | -61°59′50.8″ | 0.78±0.05 | 15.3±0.4 | 10.3±0.6 | 79±5 | 1.56 | <0.17 | 6 |
| G296.7256-01.0382 | – | – | <0.27 | – | – | – | – | – | |
| G297.1390-01.3510 | 11$^h$58′54.95″ | -63°37′47.2″ | 1.34±0.16 | 11.5±1.6 | 10.1±1.5 | 168±44 | 0.66 | <-0.69 | 6 |
| G298.2620+00.7394 | 12$^h$11′47.79″ | -61°46′20.5″ | 0.04±0.01 | – | – | – | 1.87 | 0.94±0.13 | 4 |
| G299.5265+00.1478 | – | – | <0.25 | – | – | – | – | – | – |
| G300.3412-00.2190 | – | – | <0.30 | – | – | – | – | – | – |
| G300.7221+01.2007 | 12$^h$32′50.02″ | -61°35′24.7″ | 0.38±0.02 | 19.3±0.8 | 6.5±0.4 | 25±2 | 2.48 | <0.36 | 6 |
| G301.0130+01.1153 | 12$^h$35′14.21″ | -61°41′46.7″ | 0.21±0.03 | – | – | – | 1.85 | <1.46 | 6 |
| G301.1726+01.0034 | – | – | <0.27 | – | – | – | – | – | – |
| G304.3674-00.3359A | – | – | <0.45 | – | – | – | – | – | – |
| G304.7592-00.6299 | 13$^h$07′47.14″ | -63°26′37.3″ | 0.63±0.02 | – | – | – | 1.90 | <0.05 | 6 |
| G304.8872+00.6356 | – | – | <0.27 | – | – | – | – | – | – |
| G305.5610+00.0124 | 13$^h$14′26.11″ | -62°44′29.6″ | 0.07±0.02 | 12.4±0.7 | 12.2±0.5 | 129±10 | 1.89 | 1.24±0.43 | 4 |
| G306.1160+00.1386A | – | – | <0.21 | – | – | – | – | – | – |
| G306.1160+00.1386B | – | – | <0.21 | – | – | – | – | – | – |
| G309.4230-00.6208 | – | – | <0.24 | – | – | – | – | – | – |
| G309.9796+00.5496 | 13$^h$51′02.74″ | -61°30′13.7″ | 0.80±0.03 | 7.4±0.3 | 5.6±0.3 | 74±6 | 0.42 | <-0.23 | 6 |
| G310.0135+00.3892 | 13$^h$51′37.61″ | -61°39′07.6″ | 0.76±0.02 | – | – | – | 1.71 | 0.31±0.01 | 4 |
| G310.1420+00.7583A | 13$^h$51′58.44″ | -61°15′41.5″ | 11.0±0.7 | 13.4±1.0 | 3.2±1.0 | 80±3 | 1.24 | -0.14±0.03 | 4 |
| G311.0593-00.3349 | – | – | <0.72 | – | – | – | – | – | – |
| G311.2292-00.0315 | – | – | <0.81 | – | – | – | – | – | – |
| G311.5671+00.3189 | – | – | <0.39 | – | – | – | – | – | – |
| G311.9799-00.9527 | – | – | <0.14 | – | – | – | – | – | – |
| G312.0963-00.2356 | – | – | <1.26 | – | – | – | – | – | – |
| G314.3197+00.1125 | 14$^h$26′26.47″ | -60°38′35.4″ | 0.41±0.03 | 19±3 | 7±2 | 166±6 | 4.14 | <0.14 | 6 |
| G316.5871-00.8086 | – | – | <0.18 | – | – | – | – | – | – |
| G318.9480-00.1969A | 15$^h$00′55.29″ | -58°58′52.5″ | 0.31±0.03 | – | – | – | 0.18 | -0.27±0.18 | 4 |
| G319.8366-00.1963 | – | – | <0.21 | – | – | – | – | – | – |
| G320.2437-00.5619 | – | – | <0.54 | – | – | – | – | – | – |
| G321.3824-00.2861 | – | – | <0.54 | – | – | – | – | – | – |
| G326.4755+00.6947 | – | – | <1.20 | – | – | – | – | – | – |
| G327.1192+00.5103 | 15$^h$47′32.79″ | -53°52′39.8″ | 0.30±0.04 | – | – | – | 0.51 | 0.17±0.09 | 4 |
| G327.3941+00.1970 | – | – | <0.14 | – | – | – | – | – | – |
| G327.6184-00.1109 | – | – | <0.45 | – | – | – | – | – | – |
| G328.2523-00.5320A | 15$^h$57′59.64″ | -53°58′01.0″ | 0.92±0.03 | – | – | – | 1.70 | <-0.13 | 6 |
| G328.2658+00.5316 | – | – | <1.20 | – | – | – | – | – | – |
| G328.3442-00.4629 | 15$^h$58′09.61″ | -53°51′18.3″ | 0.93±0.03 | 12.2±0.1 | 10.9±0.1 | 54±3 | 0.13 | <0.73 | 6 |
| G328.5487+00.2717 | – | – | <0.20 | – | – | – | – | – | – |
| G328.5657+00.4233 | – | – | <0.26 | – | – | – | – | – | – |
| G328.9842-00.4361 | 16$^h$01′18.99″ | -53°25′03.6″ | 0.63±0.03 | 10.4±0.1 | 7.8±0.1 | 164±8 | 1.31 | <0.16 | 6 |

Continued on next page

Table A1 – continued from previous page

| | | | | | | | | | |
|---|---|---|---|---|---|---|---|---|---|
| G329.0663-00.3081 | 16$^h$01′09.98″ | -53°16′01.2″ | 0.33±0.02 | 11.3±0.4 | 6.5±0.4 | 42±5 | 1.19 | <0.32 | 6 |
| G329.3402-00.6436 | – | – | <0.27 | – | – | – | – | – | – |
| G329.6098+00.1139 | – | – | <0.72 | – | – | – | – | – | – |
| G330.0699+01.0639 | – | – | <1.50 | – | – | – | – | – | – |
| G330.2923+00.0010A | – | – | <0.63 | – | – | – | – | – | – |
| G331.0890+00.0163A | – | – | <0.30 | – | – | – | – | – | – |
| G331.5651+00.2883 | – | – | <1.80 | – | – | – | – | – | – |
| G331.7953-00.0979 | – | – | <0.90 | – | – | – | – | – | – |
| G333.1153+00.0950 | – | – | <1.38 | – | – | – | – | – | – |
| G333.7608-00.2253 | – | – | <0.80 | – | – | – | – | – | – |
| G334.7302+00.0052 | 16$^h$26′04.69″ | -49°08′41.9″ | 0.12±0.01 | – | – | – | 0.41 | <1.42 | 6 |
| G335.0611-00.4261A | 16$^h$29′23.05″ | -49°12′27.5″ | 1.18±0.16 | 13.0±2.0 | 10±2 | 22±34 | 0.85 | – | – |
| G337.0963-00.9291 | – | – | <0.72 | – | – | – | – | – | – |
| G338.2717+00.5211A | – | – | <0.24 | – | – | – | – | – | – |
| G338.2717+00.5211B | – | – | <0.24 | – | – | – | – | – | – |
| G338.2801+00.5419A | 16$^h$38′09.00″ | -46°10′60.0″ | 0.43±0.08 | – | – | – | 4.15 | <-0.02 | 6 |
| G338.4712+00.2871 | 16$^h$39′58.59″ | -46°12′37.0″ | 0.80±0.03 | – | – | – | 3.27 | <0.19 | 6 |
| G339.3316+00.0964 | 16$^h$44′04.27″ | -45°41′26.6″ | 2.12±0.04 | – | – | – | 1.35 | <-0.72 | 6 |
| G339.6221-00.1209 | 16$^h$46′05.82″ | -45°36′42.9″ | 4.71±0.06 | 10.9±0.2 | 8.7±0.2 | 64±3 | 2.04 | -0.64±0.06 | 1,4 |
| G339.9267-00.0837 | – | – | <0.75 | – | – | – | – | – | – |
| G340.7455-01.0021 | 16$^h$54′03.94″ | -45°18′48.9″ | 1.63±0.05 | – | – | – | 1.60 | – | – |
| G342.9583-00.3180 | – | – | <0.33 | – | – | – | – | – | – |
| G343.1261-00.0623 | 16$^h$58′17.12″ | -42°52′06.3″ | 33.2±2.5 | 24.1±2.2 | 3.7±1.2 | 161±2 | 1.19 | -0.17±0.09 | 4,6 |
| G343.5213-00.5171 | 17$^h$01′33.88″ | -42°50′18.7″ | 0.67±0.05 | 7.3±0.4 | 7.2±0.4 | 4±164 | 2.02 | 0.02±0.01 | 4 |
| G343.8354-00.1058 | – | – | <0.90 | – | – | – | – | – | – |
| G344.6608+00.3401 | – | – | <0.25 | – | – | – | – | – | – |
| G344.8746+01.4347 | – | – | <1.50 | – | – | – | – | – | – |
| G344.8889+01.4349 | – | – | <1.20 | – | – | – | – | – | – |
| G345.2619-00.4188A | 17$^h$06′50.47″ | -41°23′45.5″ | 0.27±0.03 | 9.5±1.1 | 3.2±1.4 | 131±5 | 1.20 | <0.57 | 6 |
| G345.4938+01.4677 | 16$^h$59′41.51″ | -40°03′43.4″ | 51.8±0.8 | 20.5±0.4 | 6.6±0.2 | 101±5 | 1.15 | -0.30±0.10 | 13 |
| G345.5043+00.3480 | 17$^h$04′22.82″ | -40°44′22.2″ | 1.77±0.07 | – | – | – | 1.42 | <-0.13 | 14 |
| G345.7172+00.8166A | 17$^h$03′06.25″ | -40°17′08.7″ | 0.68±0.03 | 15.2±0.6 | 8.2±0.4 | 24±3 | 1.72 | <0.01 | 6 |
| G345.9561+00.6123 | 17$^h$04′42.90″ | -40°13′12.2″ | <0.09 | – | – | – | – | – | – |
| G347.0775-00.3927 | – | – | <0.33 | – | – | – | – | – | – |
| G348.6132-00.9096 | 17$^h$19′14.17″ | -38°58′21.2″ | 5.50±0.15 | – | – | – | 1.49 | <-0.33 | 6 |
| G348.6491+00.0225B | 17$^h$15′26.45″ | -38°24′15.9″ | 1.94±0.19 | – | – | – | 1.12 | – | – |
| G349.1469-00.9765 | – | – | < 0.27 | – | – | – | – | – | – |
| G349.6433-01.0957A | 17$^h$23′01.03″ | -38:13′50.1″ | 0.50±0.03 | 19.4±0.2 | 7.4±0.2 | 162±3 | 2.45 | – | – |

[1]Rodríguez et al. (2008)
[2]van der Walt et al. (2003)
[3]Zapata et al. (2006)
[4]Purser et al. (2016)
[5]Becker et al. (1994)
[6]Purcell et al. (2013)
[7]Menten & van der Tak (2004)
[8]Rosero et al. (2016)
[9]Purser et al. (2021)
[10]Sánchez-Monge et al. (2011)
[11]Carral et al. (1999)
[12]Felli et al. (1998)
[13]Guzmán et al. (2016)
[14]Garay et al. (2006)
[15]Medina et al. (2019)